\documentstyle[12pt]{article}
\begin{document}
\begin{titlepage}

\centerline{\bf Lattice $\varphi^4$ theory  of finite-size effects 
above the upper critical dimension }

\vspace*{0.5cm}
\centerline{X. S. Chen $^{1,2}$ and V. Dohm $^{1}$}
\vspace{0.4cm} 
\centerline{$^1$Institut f\"ur Theoretische Physik, Technische Hochschule 
Aachen,}
\centerline{D-52056 Aachen, Germany}     
\vspace{0.4cm}
\centerline{$^{2}$ Institute of Particle Physics, Hua-Zhong Normal University,}
\centerline{Wuhan 430079, China}
\vspace{0.4cm}
\centerline{(17 September 1998)} 
\vspace{0.4cm}

\begin{abstract}
We present a perturbative calculation of finite-size effects
near $T_c$ of the 
$\varphi^4$ lattice model in a $d$-dimensional cubic geometry of size $L$ with 
periodic boundary 
conditions for $d > 4$. The structural differences 
between the $\varphi^4$ lattice theory and the $\varphi^4$  field 
theory found previously in the spherical
limit are shown to  exist also for a finite number of components
of the order parameter. The two-variable finite-size scaling functions of the 
field theory are nonuniversal whereas those of the  lattice theory
are  independent of the nonuniversal model parameters.
One-loop results for finite-size scaling functions are derived. Their structure
disagrees with the single-variable scaling form of the lowest-mode 
approximation for any finite $\xi/L$ where $\xi$ is the bulk correlation
length. At $T_c$, the large-$L$ behavior becomes lowest-mode like for
the lattice model but not for the field-theoretic model. 
Characteristic temperatures close to $T_c$ of the 
lattice model, such as $T_{max}(L)$ of the maximum of the susceptibility 
$\chi$, 
are found to scale asymptotically as $T_c - T_{max}(L) \sim L^{-d/2}$, 
in agreement with previous Monte Carlo (MC) data for the five-dimensional 
Ising model. We also predict $\chi_{max}
\sim L^{d/2}$ asymptotically.
On a quantitative level, the  asymptotic amplitudes of this 
large -$L$ behavior close to $T_c$ have not been observed in previous 
MC simulations at $d = 5$ because of nonnegligible finite-size terms 
$\sim  L^{(4-d)/2}$ caused by the inhomogeneous modes. These terms identify 
the possible origin of a significant discrepancy between the lowest-mode 
approximation and previous MC data. MC data of larger systems would be 
desirable for testing the magnitude of the $L^{(4-d)/2}$ and $L^{4-d}$ terms
predicted by our theory.
\end{abstract}

\noindent
PACS numbers: 05.70.Jk, 64.60.-i, 75.40.Mg
\end{titlepage}

\newpage

\subsection*{1. Introduction}

A detailed understanding of the range of applicability of the
$\varphi^4$ field theory in $d$ dimensions is of fundamental interest
to statistical and elementary particle physics \cite{ZJ}.
The perturbative treatment of the critical behavior of the $\varphi^4$
field theory in $d \leq 4$
dimensions is known to be nontrivial because of the problem of infrared
divergences. This problem is solved by the renormalization-group theory 
\cite{ZJ,BGZJ}.
Above four dimensions where the critical behavior is mean-field like, 
no infrared 
problems of perturbation theory arise and no necessity exists for invoking
the renormalization group. Thus the $\varphi^4$ theory above four dimensions
appears to be free of essential problems.

This is true, however, only for infinite systems. For the $\varphi^4$
field theory of confined systems above four dimensions \cite{ZJ,B1}
there are several aspects of general interest $-$ such as the nature of the
fundamental reference lengths, the range of validity of universal finite-size
scaling, the
relevance of inhomogeneous fluctuations, and the significance of 
lattice effects 
$-$ that have remained unresolved until recently. 
These  issues have turned out [4-6] to be closely related to the 
longstanding problem 
regarding the verification of earlier phenomenological
\cite{BNP,B2} and analytical \cite{B1} 
predictions  for $d > 4$
and regarding the various attempts to test these predictions by means of 
Monte Carlo (MC) simulations for the five-dimensional
Ising model [7-12]. 
Clarifying these issues is of substantial interest for a better understanding
of finite-size effects and of the concept of finite-size scaling [13-18], 
not only for $d > 4$ but also for the limit $d \rightarrow 4$.

Recently [4-6] we have shown, on the basis of exact results in
the limit $n \rightarrow \infty$ of the $O(n)$ symmetric $\varphi^4$
theory, that finite-size effects for $d > 4$, for cubic geometry and 
periodic boundary conditions, are  more complicated and less universal
than predicted previously  and that therefore the previous
analyses of MC data were not conclusive. In particular we have found that
lattice effects and inhomogeneous fluctuations of the order parameter 
play an unexpectedly import role and that two reference lengths must be 
employed in a finite-size scaling description.  So far, however, no
direct justification was given for our conclusions to be valid also for
the more relevant  case of lattice systems with a {\it{finite}} number
$n$ of components of the order parameter.

It is the purpose of the present paper to provide this justification. 
We shall present a perturbative treatment of a $\varphi^4$ {\it{lattice}}
model  in one-loop order that leads to quantitative predictions of 
asymptotic finite-size effects for 
$d > 4$ and $n = 1$. 
We shall show that the previous arguments \cite{CD1} demonstrating
the necessity of using two scaling variables (rather than a single scaling
variable) remain valid also for finite $n$ for both the field-theoretic and
the lattice model. We also confirm that the finite-size effects of 
the $\varphi^4$ lattice model differ fundamentally from those of the 
$\varphi^4$ field theory for general $n$. This implies that the 
Landau-Ginzburg-Wilson continuum Hamiltonian for an $n$ component order
parameter does not correctly describe the finite-size effects of spin models
on lattices with periodic boundary conditions above the upper critical 
dimension.

More specifically, we study the case of  cubic geometry (volume $L^d$) 
with periodic boundary
conditions and calculate the asymptotic finite-size scaling form of the
order-parameter distribution function $P(\Phi)$  where $\Phi$ is the
spatial average of the fluctuating local order parameter $\varphi$. From
$P(\Phi)$ we derive the asymptotic finite-size scaling functions 
of the susceptibilty, 
of the order parameter 
and of the Binder cumulant \cite{14}.
{\it{Two}} scaling variables 
\begin{equation}
x = t(L/\xi_0)^2 \quad , \quad t = (T-T_c)/T_c
\end{equation}
and
\begin{equation}
y = (L/l_0)^{4-d}
\end{equation}
are needed where $\xi_0$ and $l_0$ are  reference lengths related to
the bulk correlation length $\xi$, similar to the case 
$n \rightarrow \infty$ [4-6].
$\xi_0$ is the amplitude of $\xi$
for $T > T_c$ at vanishing external field $h$ whereas $l_0$ is 
(proportional to) the amplitude of $\xi$ at $T = T_c$ for small $h$ \cite{CD2}.
The second length $l_0$ is connected with the four-point coupling $u_0$
according to $l_0 \sim u_0^{1/(d-4)}$. As an alternative choice of scaling
variables we also employ $w$ and $y$ where
\begin{equation}
w \; =\; x y^{-1/2} \; = \; t (L/\tilde{\ell})^{d/2}
\end{equation}
with the reference length
\begin{equation}
\tilde{\ell} \; = \; l_0 (\xi_0/l_0)^{4/d} \quad .
\end{equation}
In addition to the lengths $\xi_0$, $l_0$ and $L$ there is the microscopic
length $\tilde{a}$ or $\Lambda^{-1}$, i.e., the lattice spacing of the 
lattice model or the inverse cutoff of the field-theoretic model. 
For short-range 
interactions, $\xi_0$ and $l_0$ are expected to be of ${\it O}(\tilde{a})$ or
${\it O}(\Lambda^{-1})$. Our scaling functions presented in Section 4 are 
valid in the asymptotic
range $L \gg \tilde{a}$, $\xi \gg \tilde{a}$ or $L\Lambda \gg 1$, 
$\xi \Lambda \gg 1$.

The role of the length $l_0$ is twofold. Since the dangerous  irrelevant 
character
of $u_0$ \cite{F2,PF} exists  already at the mean-field level, the length $l_0$
appears via $\tilde{\ell}$ in  the variable 
$w \; \sim \; u_0^{-1/2} t L^{d/2}$ already at the level of the lowest-mode
approximation which takes only homogeneous fluctuations into account 
\cite{B1}. The second important role of $l_0$ originates from
$u_0$ being the coupling of the {\it{inhomogeneous}} higher  modes. 
Here $u_0$ does not have a dangerous irrelevant character. In fact,
these higher modes are {\it{relevant}} for $d > 4$ as has been 
demonstrated for $n \rightarrow \infty$
\cite{CD1}  and will be shown in the present paper to be valid for
general $n$, contrary to opposite statements in the literature [1,3]. 
The length $l_0$ sets the length scale of the finite-size
effects arising from these modes.

Both scaling variables $x$ and $y$ or $w$ and $y$ must be employed in 
general, i.e., at any finite
value of $\xi/L < \infty$ in the entire asymptotic $L^{-1} - \xi^{-1}$ plane 
(Fig.1) to provide a complete description of asymptotic finite-size effects 
of the $\varphi^4$ theory. Our description is
consistent with the general scaling 
structure in terms of $tL^2$ and $u_0 L^{4-d}$ proposed by Privman and 
Fisher \cite{PF} but inconsistent with the 
reduced structure proposed by Binder et al. [7] and with the lowest-mode 
approximation of Br\'{e}zin and Zinn-Justin [3] in terms of a single scaling 
variable $tL^{d/2}$ equivalent to $w$.
We find that it is only the region between the curved dotted lines in Fig.1 
where the single-variable scaling forms of Refs. [3] and [7] are justifiable
for the lattice model, but not for the field-theoretic model. The region 
between the curved lines corresponds to
the special case $\xi/L \rightarrow \infty$
in the limit $L \rightarrow \infty$ and $|t| \rightarrow 0$ 
at finite $w$ where the large - $L$
behavior becomes lowest-mode like for the lattice model. 
For $t = 0$ ($w=0$) this was found previously [4-6] for 
the case of the susceptibility  and of the Binder cumulant, as conjectured in
Ref. \cite{15}. 
As a consequence we shall show that characteristic temperatures, in the sense 
of ''pseudocritical'' temperatures \cite{F} such as $T_{max}(L)$
of the maxima of the susceptibility or the ''effective critical temperature''
\cite{RNB} $T_c(L)$ where the magnetization has its maximum slope, scale
asymptotically as $T_c - T_{max}(L) \sim L^{-d/2}$ or 
$T_c - T_c(L) \sim L^{-d/2}$ for the $\varphi^4$ lattice model, 
in agreement with previous Monte Carlo (MC) data \cite{RNB}.
Correspondingly we predict $\chi_{max}\sim L^{d/2}$ asymptotically 
for the lattice model.

On a quantitative level, our theory predicts that the large  -$L$ behavior 
close to $T_c$ is strongly affected by nonnegligible 
finite-size terms $\sim L^{(4-d)/2}$ caused by the higher (inhomogeneous)
modes, as demonstrated  
recently for the Binder cumulant at $T_c$ \cite{CD3}. In the analysis of
Ref. [9] the observed ''slow approach to the scaling limit'' was considered
to be the most significant discrepancy between the lowest-mode prediction
\cite{B1} and the MC data. Our theory identifies the terms 
$\sim L^{(4-d)/2}$ as the possible origin of
this discrepancy. We also show 
that, for the same reason, the successful method of determining bulk
$T_c$ via the intersection point of the Binder cumulant \cite{14}
is not accurately applicable to finite spin models of small size 
in $d = 5$ dimensions, as  demonstrated in Ref. [6]. New MC simulations over
a larger range of $L$ would be desirable for testing the predictions of our
theory regarding the magnitude of the $L^{(4-d)/2}$ and $L^{4-d}$ terms.

In Section II we derive some of the bulk properties of the lattice model
for $n = 1$ above four dimensions in one-loop order. In particular we 
identify the amplitudes of the correlation length at $h = 0$
for $T > T_c$ as well as at $T = T_c$ for $h \neq 0$. 
In Section III we derive the effective Hamiltonian
and the order-parameter distribution function in one-loop order. Applications
to the asymptotic (large $L \gg \tilde{a}$, small $|t| \ll 1$) finite-size 
scaling functions 
and predictions of the lattice model for $d = 5$  are presented and 
discussed in Section IV. Results for the field-theoretic model are briefly
presented in Section V. We summarize the results of our paper in Section VI.

\subsection*{2. Lattice model: Bulk properties for ${\bf{d > 4}}$}

We  consider a $\varphi^4$ lattice
Hamiltonian $H$ for the one-component variables $\varphi_i$
with  \mbox{$ - \infty \leq 
\varphi_{i} \leq \infty$} 
on the lattice
points ${\bf{x}}_i$ of a simple-cubic lattice in a cube with volume $V = L^d$
and with periodic boundary conditions. We assume 

\begin{equation}\label{Hhat}
H \; = \;  \tilde{a}^d \; 
\left \{ \sum\limits_i [ \frac{r_0}{2}
\varphi^2_i \; + \; u_0 (\varphi_i^2)^2 - h \varphi_i]\; + \;
\sum\limits_{i,j} \frac{1}{2 \tilde{a}^2} J_{ij} 
(\varphi_i - \varphi_j)^2 \right \}
\end{equation}
where $\tilde{a}$ is the lattice spacing.
The couplings $J_{ij}$ are dimensionless quantities.  The  variables 
$\varphi_j$ have the Fourier representation
\begin{equation}
\varphi_j \; = \; \frac{1}{L^{d}} \; \sum\limits_{\bf{k}} \; 
e^{i {\bf{k}} \cdot {\bf{x}}_j} \hat{\varphi}_{\bf{k}} \quad .
\end{equation}
In terms of the Fourier components
\begin{equation}\label{var}
\hat{\varphi}_{\bf{k}} \; = \; \tilde{a}^d \sum\limits_j \; 
e^{-i{\bf{k \cdot x}}_j} \varphi_j
\end{equation}
the Hamiltonian $H$ reads
\begin{eqnarray}\label{Hhat=}
H & = &  L^{-d} \;  \sum\limits_{\bf{k}}
\frac{1}{2}\; [ r_0 \; + \; \hat{J}_{\bf{k}} ]
\hat{\varphi}_{\bf{k}} \; \hat{\varphi}_{-{\bf{k}}} 
\; - \; h \hat{\varphi}_{{\bf{0}}} \nonumber \\[0.4cm]
&&  +  \; u_0 L^{-3d} \sum\limits_{{\bf{k k' k''}}} \;
(\hat{\varphi}_{{\bf{k}}} \; \hat{\varphi}_{\bf{{k'}}}) 
(\hat{\varphi}_{{\bf{k''}}} \; \hat{\varphi}_{{\bf{-k-k'-k''}}}) 
\end{eqnarray}
where
\begin{equation}\label{D}
\hat{J}_{\bf{k}} \; = \;
\frac{2}{\tilde{a}^2} \; [J(0) \; - \; J({\bf{k}})] 
\end{equation}
with
\begin{equation}\label{J}
J({\bf{k}}) \; = \;(\tilde{a}/L)^{d} \; \sum\limits_{i,j} \; J_{ij} 
e^{-i{\bf{k}}\cdot ({\bf{x}}_i-{\bf{x}}_j)} \; .
\end{equation}
The   summation   $\; \; \sum_{\bf{k}} \; \; $   runs over discrete   
$ \; {\bf{k}}\;  $  vectors with  components $\; k_j \;  =
2 \pi m_j / L,  \; m_j = 0 , \pm 1, \pm 2, \cdots, \; 
\; j = 1, 2, \cdots, d $  with a finite cutoff 
{\mbox {$-\Lambda \equiv - \pi/\tilde{a} \leq k_j $}} 
$ < \pi/\tilde{a} \equiv \Lambda$. 
We assume a finite-range pair interaction $J_{ij}$ such
that its Fourier transform  has the small ${\bf{k}}$
behavior
\begin{equation}
\hat{J}
_{\bf{k}} \; = \; J_0 {\bf{k}}^2 \quad + \quad O (k_i^2 k_j^2)
\end{equation}
with
\begin{equation}\label{J0}
J_0 \; = \; \frac{1}{d} (\tilde{a}/L)^d \; \sum\limits_{i,j}
\left ( J_{ij}/\tilde{a}^2 \right )  \left ( {\bf{x}}_i \; - \; {\bf{x}}_j 
\right )^2 \quad .
\end{equation}

The complete information on thermodynamic properties is contained in the
Gibbs free energy  per unit volume (in units of $k_B T$)
\begin{equation}\label{V}
f \; = \; - 
\frac{1}
{L^d} \ln \int {\cal{D}} \varphi
\exp (- H) 
\end{equation}
where the symbol $\int {\cal{D}}\varphi$ is the usual abbreviation for the
multiple integral over the real and imaginary parts of (the finite
number of) the Fourier components $ \varphi_{\bf{k}}$.

Recently we have found \cite{CD2}
in the large-$n$ limit that the two reference lengths of
the finite-size scaling functions for $d > 4$ are determined by the
two {\it{bulk}} correlation-length amplitudes $\xi_0$
at $h = 0$ for $T > T_c$ and $l_0$ at $T = T_c$ for small $h$.
We shall see that this property remains valid, as far as $\xi_0$ is concerned,
 also for $n = 1$ in 
one-loop order. With regard to $l_0$, an additional $n$-dependent prefactor
arises
that is $1$ in the large-$n$ limit and $3^{-1/2} 2^{-1/3}$ for $n = 1$
in one-loop order (see Eq. (28) below).
For this purpose we need to identify the amplitudes of the bulk 
correlation length for $d > 4$. We also calculate the bulk amplitudes of the
order parameter $M_b$ and of  the susceptibilities $\chi^+_b$
and $\chi^-_b$ above and below $T_c$ as reference quantities 
for the corresponding finite-size effects.

All of our calculations are carried out at finite cutoff $\Lambda$ 
(finite lattice spacing $\tilde{a}$).
First we derive the asymptotic form of the bulk susceptibility $\chi_b$
at $h = 0$ above and below $T_c$ as well as
at 
$T = T_c$ for small $h$.
The bulk Gibbs free
energy density is denoted by $f_b = \lim\limits_{L \rightarrow \infty} f$.
In terms of the bulk order parameter 
\begin{equation}
M_b \; = \; - \lim\limits_{L \rightarrow \infty} \; 
\partial f/\partial h \quad 
\end{equation}
the bulk Helmholtz free
energy density  
$\Gamma_b \; = \; f_b + M_b h$ 
reads in one-loop order 
\begin{equation}
\Gamma_b (r_0, M_b) \; = \; \frac{1}{2} r_0 M^2_b + u_0 M^4_b
\; + \; \frac{1}{2} \int\limits_{\bf{k}} \ln
(r_0 + 12 u_0 M^2_b + \hat{J}_{\bf{k}}) 
\end{equation}
where $\int_{\bf{k}}$ stands for $(2\pi)^{-d} \int d^dk$
with $|k_i| \leq \Lambda$. Above $T_c$ for $h = 0$, 
the inverse bulk  susceptibility $(\chi^+_b)^{-1}$ is
\begin{equation}
(\chi^+_b)^{-1} \; = \; (\partial^2 \Gamma_b/\partial M^2_b)_{M_b = 0} \; = \;
r_0 \; + \; 12 u_0 \int\limits_{\bf{k}} (r_0 + \hat{J}_{\bf{k}})^{-1}
\; + \; O(u^2_0)
\end{equation}
which determines the critical value $r_{0c}$ of $r_0$ as
\begin{equation}
r_{0c} \; = \; - \; 12u_0 \; \int\limits_{\bf{k}} \; \hat{J}^{-1}_{\bf{k}}
\; + \; O(u^2_0) \quad .
\end{equation}
Thus we rewrite $(\chi^+_b)^{-1}$ above $T_c$ in terms of $r_0 - r_{0c}$
as 
\begin{equation}
(\chi^+_b)^{-1} \; = \; (r_0 - r_{0c}) 
\left[ 1 - 12u_0 
\int\limits_{\bf{k}}  \hat{J}_{\bf{k}}^{-2}
\right] \; + \; O \; (u^2_0) 
\end{equation}
where
\begin{equation}
r_0 - r_{0c} \; = \; a_0 t, \quad t \; = \; (T - T_c)/T_c \quad .
\end{equation}
Note that the integral in Eq. (18) exists only for $d > 4$ and only for 
finite cutoff. 
The spontaneous value $M_s$ of the bulk order parameter for $h \rightarrow 0$
below $T_c$ is determined by $\partial \Gamma_b / \partial M_b = 0$. 
This yields for $d > 4$
\begin{equation}
M^2_s \, = \; (4u_0)^{-1} (r_{0c} - r_0)
\left[ 1 + 24 u_0 \int\limits_{\bf{k}} \hat{J}^{-2}_{\bf{k}} 
\right]
+ O(u^2_0) \quad .
\end{equation}
The inverse susceptibility $(\chi^-_b)^{-1}$ for $h \rightarrow 0$ below $T_c$ is in one-loop
order
\begin{equation}
(\chi^{-}_b)^{-1} \; = \; \left ( \frac{\partial^2 \Gamma}{\partial M^2_b}
\right )_{M_b = M_s} \; = \; 2 \left(r_{0c} - r_{0}
\right)\left[1 - 12 u_0 \int\limits_{\bf{k}} \hat{J}^{-2}_{\bf{k}} 
\right]+ O(u^2_0)\quad .
\end{equation}
From the equation of state at $T_c$
\begin{equation}
\frac
{\partial \Gamma_b}{\partial M_b}
\; = \; h \; = \; 4 u_0 M^3_b \;
\left[
1 - 36 u_0 \; \int\limits_{\bf{k}} \;
\hat{J}^{-2}_{\bf{k}}
\right]
\end{equation}
we obtain the $h$ dependence of the inverse bulk susceptibility 
$\chi_c^{-1}$ at $T_c$
as 
\begin{equation}
\chi_c^{-1} \; = \;
\frac
{\partial^2 \Gamma_b}{\partial M_b^2}
\; = \;  3 h^{2/3}
\left\{ 4 u_0 \left[ 1 - 36 u_0 \int\limits_{\bf{k}} \hat{J}_{\bf{k}}^{-2} 
\right] 
\right\}
^{1/3} \; + \; O(u^3_0) \; .
\end{equation}
For $T \geq T_c$,
the bulk susceptibility at finite wave vector ${\bf{q}}$
\begin{equation}\label{W}
\chi_b ({\bf{q}}) \; = \;\lim\limits_{L \rightarrow \infty} 
\frac
{\tilde{a}^{2d}}{L^d} 
\sum\limits_{i,j} <  \varphi_i \; 
\varphi_j >  \; e^{-i{\bf{q}} \cdot ({\bf{x}}_i - {\bf{x}}_j)} \quad 
\end{equation}
has the one-loop form (for both $h = 0 $ and $h \not= 0$)
\begin{equation}
\chi_b({\bf{q}})^{-1} \; = \; \chi_b ({\bf{0}})^{-1} \; + \; 
\hat{J}_{\bf{q}} 
\left[
1 + O(u^2_0)
\right] \quad .
\end{equation}
Thus the square of the bulk correlation length for $T \geq T_c$ is given by
\begin{equation}\label{chide}
\xi^2 \; = \; \chi_b \; ({\bf{0}})
\left [ \partial \chi_b
({\bf{k}})^{-1}/ \partial {\bf{k}}^2 \right ]_{{\bf{k}} = {\bf{0}}} \; = \;
J_0 \; \chi_b ({\bf{0}}) \; + \; O(u^2_0) \;  .
\end{equation}
Substituting Eqs. (18) and (23) into Eqs. (25) and (26) 
yields the asymptotic form for $d > 4$
\begin{equation}
\xi \; = \; \xi_0 t^{-1/2}\quad , \quad t > 0 \; , \;  h = 0\; ,
\end{equation}
and
\begin{equation}
\xi \; = \; 3^{-1/2} 2^{-1/3} l_0 (h^2 l_0^{d+2} J_0^{-1})^{-1/6}
\; , \; t = 0 \; , \; h \not= 0
\end{equation}
with the lengths
\begin{eqnarray}
\xi_0 \; & = & \; a_0^{-1/2} \; J_0^{1/2}
\left[ 1 + 12 u_0 \; \int\limits_{\bf{k}} \; \hat{J}^{-2}_{\bf{k}} 
\right]^{1/2}
\quad + \quad O(u^2_0) 
\end{eqnarray}
and
\begin{eqnarray}
l_0 \; & = & \; \left\{u_0 J^{-2}_0 
\left [ 1 \; + 36 u_0  \;  \int\limits_{\bf{k}}
\hat{J}_{\bf{k}}^{-2}
\right]^{-1} 
\right \}^{1/(d-4)}.
\end{eqnarray}
These lengths will appear also in the finite-size scaling
functions in Sect. IV.
We see that for $d > 4$ the fluctuations (that enter via the one-loop
integrals) only modify the amplitudes but do not change the mean-field $t$
and $h$ dependence. The ''dangerous'' $u_0$ dependence \cite{F2} of $\xi$
at $T_c$, Eqs. (28) and (30),
is clearly seen in $l_0 \sim u_0^{1/(d-4)}$.
We note that both $\xi_0$ and $l_0$ are cutoff dependent via $\int_{\bf{k}} 
\hat{J}_{\bf{k}}^{-2}$. In rewriting $[1 - 12 u_0 \int_{\bf{k}}
\hat{J}^{-2}_{\bf{k}}]^{-1/2}$ as $[1 + 12 u_0 \int_{\bf{k}}
\hat{J}^{-2}_{\bf{k}}]^{1/2}$ in $\xi_0$ (and similarly in $l_0$) 
we have been guided by the resummed forms of $\xi_0$ and $l_0$ in the 
limit $n \rightarrow \infty$ (Eqs. (141) and (142) in Ref. \cite{CD1}).

Corresponding results can be derived for the continuum $\varphi^4$
Hamiltonian (see Eq. (62) below) with periodic boundary conditions, similar 
to the case $n \rightarrow \infty$ 
studied previously \cite{CD1}.
This amounts essentially to replacing $\hat{J}_{\bf{k}}$ by ${\bf{k}}^2$
in the equations given above. As far as bulk properties of 
$\chi_b^+, \chi^-_b$, $\xi$ and $M_b$ are concerned, only the nonuniversal amplitudes
are modified but the $t$ and $h$ dependence remains identical for the
field-theoretic and the lattice $\varphi^4$ model. For the finite system, 
however, lattice effects become significant as we shall see in the
subsequent Sections.  For the specific heat
even the (finite) bulk value at $T_c$ turns out to be different for the
field-theoretic and the lattice model, similar to the case 
$n \rightarrow \infty$ \cite{CD1}.

\subsection*{3. Order-parameter distribution function for ${\bf{d > 4}}$}

\subsubsection*{3.1. General form in one-loop order}

A perturbation approach to finite-size effects of the $\varphi^4$
lattice model for $d > 4$ can be set up in a way similar to the
previous finite-size perturbation theory 
for $d < 4$ above and below $T_c$ \cite{EDC}.
We decompose
\begin{equation}
\varphi_j \; = \; \Phi \; + \; \sigma_j
\end{equation}
and shall derive an effective Hamiltonian $H^{eff}$ \cite{B1,EDC}
for the lowest (homogeneous) mode
\begin{equation}
\Phi \; = \; \frac{\tilde{a}}{L^d}
\sum\limits_j \; \varphi_j
\end{equation}
by a perturbative treatment of the fluctuations of the higher modes
\begin{equation}
\sigma_j \; = \; L^{-d} \sum\limits_{{\bf{k}} \neq {\bf{0}}}
\hat{\varphi}_{\bf{k}} \; e^{i{\bf{k}} \cdot {\bf{x}}_j} \quad .
\end{equation}
Correspondingly we write the lattice Hamiltonian, Eq. (5), in the form
\begin{equation}
H \; = \; H_0(\Phi) \; + \; H_1 + H_2 \quad ,
\end{equation}
with the lowest-mode Hamiltonian
\begin{equation}
H_0 (\Phi) \; = \; L^d (\frac{1}{2} r_0 \Phi^2 \; + \; u_0 \Phi^4 - h \Phi)
\quad ,
\end{equation}
the Gaussian part
\begin{equation}
H_1 \; = \; L^{-d} \; \sum\limits_{{\bf{k}}\neq {\bf{0}}}
\frac{1}{2} (\bar{r}_{0L} + \hat{J}_{\bf{k}}) \hat{\sigma}_{\bf{k}} \;
\hat{\sigma}_{-{\bf{k}}} \; 
\end{equation}
with
\begin{equation}
\bar{r}_{0L} \; = \; r_0 \; + \; 12 u_0 M^2_0 \; ,
\end{equation}
and the perturbation part
\begin{eqnarray}
H_2 \; = \; \tilde{a}^d \; \sum\limits_j 
\left[ 6 u_0 (\Phi^2 - M^2_0)
\sigma^2_j  
 + 4 u_0 \sigma^3_j \;  +  u_0 \sigma^4_j
\right] \quad .
\end{eqnarray}
Unlike the case $d < 4$, we must work, for $d > 4$, at finite cutoff.
Thus we shall incorporate the finite shift $r_{0c} \sim O(u_0)$,
Eq. (17), of the parameter $r_0$ 
whereas in the previous \cite{EDC}
dimensional regularization at infinite cutoff 
there was no $O(u_0)$ contribution to $r_{0c}$. 
Thus we define
\begin{equation}
M^2_0 \; = \; \frac{1}{Z_0^{eff}} \; \int\limits^\infty_{-\infty}
d \Phi \; \Phi^2 \; \exp (- H^{eff}_0)
\end{equation}
where now
\begin{eqnarray}
H^{eff}_0 &\; = \; & L^d 
\left[ \frac{1}{2} (r_0 - r_{0c}) \Phi^2 \; + \; u_0  \Phi^4 - h \Phi
\right] \; \; 
\end{eqnarray}
and
\begin{equation}
Z_0^{eff} \; = \;  \int\limits^\infty_{-\infty} \; 
d \Phi \exp (- H^{eff}_0) \; .
\end{equation}
contain the shifted variable $r_0 - r_{0c}$.

The partition function is decomposed as
\begin{equation}
Z = \int {\cal{D}} \varphi \; e^{-H} \; = \;
\int\limits_{-\infty}^\infty \; d \Phi \; \exp 
\left[- (H_0 + \Gamma)
\right] \quad ,
\end{equation}
where 
\begin{equation}
\Gamma (\Phi) \; = \; - \ln \int {\cal{D}} \sigma \exp 
\left[ -(H_1 + H_2)
\right]
\end{equation}
is determined by the higher modes within 
perturbation theory. We rewrite 
\begin{equation}
H_0 (\Phi) \; + \; \Gamma (\Phi) \; = \; 
\Gamma (0) + H^{eff} (\Phi)
\end{equation}
and define the order-parameter distribution function
\begin{equation}
P(\Phi) \; = \; \frac{1}{Z^{eff}}\; \exp \left[ -H^{eff} (\Phi)
\right] \; ,
\end{equation}
\begin{equation}
Z^{eff} \; = \; \int\limits_{-\infty}^\infty \; d \Phi \exp 
\left[ - H^{eff} (\Phi)   \right] \; .
\end{equation}
In a perturbation calculation with respect to $H_2$
we obtain the effective Hamiltonian in one-loop order 
\begin{equation}
H^{eff} (\Phi) \; = \; L^d 
\left[
\frac{1}{2} r_0^{eff} \Phi^2 \; + \; u^{eff}_0 \Phi^4 \; + \; O(\Phi^6)-h\Phi
\right]\quad 
\end{equation}
where 
\begin{eqnarray}
 r^{eff}_0 = r_0 - r_{0c} + 12 u_0 \left[ L^{-d}
 \sum\limits_{{\bf{k}} \neq {\bf{0}}} (r_{0L} +
 \hat{J}_{\bf{k}})^{-1} -\int_{\bf k}\hat{J}_{\bf k}^{-2}\right]\nonumber \\ 
+ 144 u^2_0 M^2_0 L^{-d} \sum\limits_{{\bf{k}} \neq {\bf{0}}} 
(r_{0L} + \hat{J}_{\bf{k}})^{-2},
\end{eqnarray}
\begin{equation}
u^{eff}_0 \; = \; u_0 - 36 u^2_0 L^{-d} \;
\sum\limits_{{\bf{k}} \neq {\bf{0}}}
(r_{0L} \; + \; \hat{J}_{\bf{k}})^{-2} \quad .
\end{equation}
In Eq. (48) we have added and subtracted $r_{0c}$ and have replaced 
$\bar{r}_{0L}$, in the $O(u_0)$ terms, by
\begin{equation}
r_{0L} \; = \; r_0 - r_{0c} \; + \; 12 u_0 M^2_0 \quad .
\end{equation}
This quantity is a positive function of $r_0 - r_{0c}$ for arbitrary 
$- \infty \leq r_0 - r_{0c} \leq \infty$
at any finite $L$.

Moments of the distribution function can now be calculated as
\begin{equation}
< \Phi^m > \; =  \; \int\limits^\infty_{-\infty}
d \Phi \; \Phi^m \; P(\Phi)
\end{equation}
and
\begin{equation}
< |\Phi|^m > \; = \;  \int\limits^\infty_{-\infty}
d \Phi \; |\Phi|^m  P(\Phi) \quad .
\end{equation}
Because of the (one-loop)
$\Phi^4$ structure of $H^{eff}$, these averages can be expressed in terms
of the well-known functions
\begin{equation}
\vartheta_m(Y) \; = \; \frac{\int\limits_0^\infty ds \; s^m \; \exp
(-\frac{1}{2}Y s^2 \; - \; s^4)}
{\int\limits_0^\infty ds  \; \exp
(-\frac{1}{2}Y s^2 \; - \; s^4)}
\end{equation}
that appear also in the finite-size theory below four dimensions
\cite{EDC,CDS}. The moments determine several thermodynamic quantities such as 
\cite{EDC} the susceptibilities
\begin{eqnarray}
\chi^+ \; & = & \; L^d < \Phi^2> \quad ,  \\[1cm]
\chi^- \; & = & \; L^d (< \Phi^2> \; - \; < |\Phi| >^2) \quad ,
\end{eqnarray}
the ''magnetization'' $ M \; = \; < |\Phi| > $ ,
and the Binder cumulant 
\begin{equation}
U \;  = \;  1 \;  - \;  \frac{1}{3} < \Phi^4>/< \Phi^2>^2 \quad .
\end{equation}
In terms of the effective parameters $r^{eff}_0$ and $u^{eff}_0$
they can be expressed in one-loop order as 
\begin{eqnarray}
\chi^+ \; & =  & \; (L^d / u_0^{eff})^{1/2} \;
\vartheta_2 \;  (Y^{eff}) \quad ,  \\[1cm]
\chi^- \; & =  & \; (L^d / u_0^{eff})^{1/2}\; 
\left[ \vartheta_2 \;  (Y^{eff}) \; - \; \vartheta_1 \;  (Y^{eff})^2
\right] \quad ,  \\[1cm]
M \; & = & \; (L^d u_0^{eff})^{-1/4} \;
\vartheta_1 \;  (Y^{eff}) \quad ,  \\[1cm]
U \; & = & \; 1 \; - \; \frac{1}{3} \; 
\vartheta_4 (Y^{eff}) / \vartheta_2 (Y^{eff})^2 \quad ,
\end{eqnarray}
with the dimensionless quantity
\begin{equation}
Y^{eff} \; = \; L^{d/2} \; r^{eff}_0 (u_0^{eff})^{-1/2}.
\end{equation}
We note that at this stage of perturbation theory these expressions do not 
yet represent a systematic expansion with respect to the coupling $u_0$,
compare Eqs. (5.19)-(5.22) of Ref. \cite{EDC}.

Corresponding formulas are obtained for the case of the Landau-Ginzburg-Wilson
continuum Hamiltonian
\begin{equation}
H \; = \; \int\limits_V d^dx \left[ \frac{1}{2}
r_0 \varphi^2 \; + \; \frac{1}{2} (\bigtriangledown \varphi)^2
\; + \; u_0 \varphi^4 \; - \; h \varphi \right]
\end{equation}
with the field $\varphi (x)$  by the replacement $\hat{J}_{\bf{k}} \rightarrow {\bf{k}}^2$
in the sums of the one-loop terms of the effective parameters $r^{eff}_0$
and $u^{eff}_0$.

A justification of the above perturbation theory can be given
in terms of the order-parameter distribution function where
all higher modes are treated in a nonperturbative way \cite{CDS}.

\subsubsection*{3.2  Asymptotic form of the effective parameters}

In order to study the asymptotic finite-size critical behavior we shall
consider the limit of $L \gg \tilde{a} $, $\xi \gg \tilde{a}$
or $L\Lambda \gg 1$, $\xi \Lambda \gg 1$. 
For this purpose 
we decompose the perturbation part of $r^{eff}_0$ and $u^{eff}_0$ 
into bulk integrals and
finite-size contributions in the following way,
\begin{eqnarray}
r_0^{eff} & =&  (r_0 - r_{0c}) \left\{ 1 - 12 u_0 \int\limits_{\bf{k}}
\left[\hat{J}_{\bf{k}} (r_{0L} + \hat{J}_{\bf{k}})
\right]^{-1} \right\} \nonumber \\
& +& 144 u^2_0 M^2_0 
\left\{ \int_{\bf{k}} (r_{0L} + \hat{J_{\bf{k}}})^{-2}
- \int\limits_{\bf{k}}
\left[\hat{J}_{\bf{k}} (r_{0L} + \hat{J}_{\bf{k}})
\right]^{-1}\right\} \nonumber \\ 
& - & 12 u_0 \Delta_1 (r_{0L}) \;  - \;  144 u^2_0 M^2_0 
\Delta_2 (r_{0L}),
\end{eqnarray}
\begin{equation}
u_0^{eff} =   u_0 - 36 u^2_0 \int\limits_{\bf{k}}
(r_{0L} + \hat{J}_{\bf{k}})^{-2}
+ 36   u^2_0   \Delta_2 (r_{0L}) \quad ,
\end{equation}
with
\begin{equation}
\Delta_m (r_{0L}) \; = \; \int\limits_{\bf{k}}
(r_{0L} + \hat{J}_{\bf{k}})^{-m} \; - \; L^{-d}
\sum\limits_{{\bf{k}} \neq {\bf{0}}}
\; (r_{0L} + \hat{J}_{\bf{k}})^{-m} \;  .
\end{equation}
In the lowest-mode approximation we would have simply
$r^{eff}_0 \; = \; r_0 \; , \; u^{eff}_0 \; = \; u_0$.
Up to this point, the determination of the effective Hamiltonian for the
field-theoretic model, Eq. (62), is still parallel to that of the lattice 
model. 
The corresponding formulas are simply obtained by the replacement
$\hat{J}_{\bf{k}} \rightarrow {\bf{k}}^2$. The crucial difference, 
however, comes from
the large-$L$ behavior of $\Delta_m$, as shown recently for the special
case $m = 1$ and $r_{0L} = 0$ \cite{CD1}.
For general $r_{0L}$ we find, for the lattice model, the cutoff-independent
large-$L$ behavior
\begin{equation}
\Delta_m (r_{0L}) \; = \; J_0^{-m} I_m (r_{0L} J^{-1}_0 L^2) \; 
L^{2m-d} + O (e^{-L/\tilde{a}})
\end{equation}
with
\begin{equation}
I_m(x) =   (2\pi)^{-2m} \int\limits_0^{\infty} dy \; y^{m-1}\;
e^{-(xy/4\pi^2)}\left[ (\pi/y)^{d/2} \; - \; K(y)^d + 1 \right]
\end{equation}
where $K(y) =  \sum\limits_{j = - \infty}^\infty \exp (-y j^2)$. 
For the field-theoretic model, however, the corresponding large-$L$ 
behavior  differs
significantly according to the cutoff dependent result
\begin{eqnarray}
 \int_{\bf{k}} (r_{0L} + {\bf{k}}^2)^{-m} &-&  L^{-d}
\sum\limits_{{\bf{k}} \neq {\bf{0}}}(r_{0L} + {\bf{k}}^2)^{-m}  
= I_m(r_{0L} L^2) L^{2m-d} \nonumber \\ 
 &+& \Lambda^{d-2m} \left\{ a_m (d, r_{0L} \Lambda^{-2})(\Lambda L)^{-2}
+ O\left[ (\Lambda L)^{-4}\right]\right\}
\end{eqnarray}
where
\begin{equation}
a_m (d,r_{0L} \Lambda^{-2})=  \frac{d}{3(2\pi)^{d-2}} 
\int\limits^\infty_0 dx  x^m \left[ \int\limits_{-1}^1
 dy  e^{-y^2x} \right]^{d-1} \exp \left[ -(1 + r_{0L}
\Lambda^{-2})   x \right] .
\end{equation}
This leads to
\begin{eqnarray}
r_0^{eff} & =&  (r_0 - r_{0c}) \left\{ 1 - 12 u_0 \int\limits_{\bf{k}}
\left[{\bf k}^2 (r_{0L} + {\bf k}^2 )
\right]^{-1} \right\} \nonumber \\
& +& 144 u^2_0 M^2_0 
\left\{ \int_{\bf{k}} (r_{0L} + {\bf k}^2)^{-2}
- \int\limits_{\bf{k}}\left[ {\bf k}^2(r_{0L} + {\bf k}^2)
\right]^{-1}\right\} \nonumber \\ 
& - & 12 u_0 \left[I_1 (r_{0L}L^2)L^{2-d}+\Lambda^{d-2}
a_1 (d,r_{0L}\Lambda^{-2})(\Lambda L)^{-2}\right] \nonumber \\
& -&   144 u^2_0 M^2_0 \left[I_2 (r_{0L}L^2)L^{4-d}+\Lambda^{d-4}
a_2 (d,r_{0L}\Lambda^{-2})(\Lambda L)^{-2}\right],
\end{eqnarray}
\begin{eqnarray}
u_0^{eff} =   u_0 &-& 36 u^2_0 \int\limits_{\bf{k}}
(r_{0L} + {\bf k}^2 )^{-2} \nonumber \\
&+& 36   u^2_0  \left[I_2 (r_{0L}L^2)L^{4-d}+\Lambda^{d-4}
a_2 (d,r_{0L}\Lambda^{-2})(\Lambda L)^{-2}\right]
\end{eqnarray}
for the field-theoretic model.
We note that for $d < 4$ the cutoff dependent terms in Eqs. (70) and (71) 
vanish in the limit $\Lambda \rightarrow \infty$. For $d > 4$, however, 
part of these terms are
divergent for $\Lambda \rightarrow \infty$ and cannot be dropped. In particular
these terms carry the important size dependence $\sim L^{-2}$ of $r_0^{eff}$ 
which is not present in Eq.(63) for the lattice model and which has been
overlooked previously \cite{ZJ,B1}.
Employing the method of dimensional regularization \cite{ZJ} in Eq. (68) 
would mean that 
the finite results for $d < 4$ at $\Lambda = \infty$ are continued 
analytically to $d > 4 $.
This would yield the same 
(cutoff-independent) result for $r_0^{eff}$  and $u_0^{eff}$ of the 
field-theoretic case as of the lattice model. 
Thus dimensional regularization would omit the important analytic $L^{-2}$ 
dependence in Eqs. (68) and (70). 
This omission, however, cannot be justified $-$ unlike the omission of an 
analytic $t$ dependence in bulk critical phenomena.
Thus the method of dimensional regularization may yield misleading results
in the finite-size field theory above the upper critical dimension.

The remaining bulk integrals in $r^{eff}_0$ and $u^{eff}_0$
have finite limits for $r_{0L} \rightarrow 0$ (large $L$, small $|t|$)
for both the lattice and field-theoretic model. Taking the limit 
$r_{0L} \rightarrow 0$ in these integrals is justified only if $|t| \ll 1$
and $(L/\tilde{a})^{-d/2}\ll 1$ or $(\Lambda L)^{-d/2}\ll 1$. This restriction
should be kept in mind when applying our asymptotic scaling functions to MC 
data of spin models of small size. The asymptotic 
expressions for the lattice model for $d > 4$ read
\begin{eqnarray}
r^{eff}_0  =  (r_0 - r_{0c})\left[ 1 - 12 u_0 \int_{\bf{k}} 
\hat{J}_{\bf{k}}^{-2}
\right] - 12 u_0 J_0^{-1} L^{2-d}  I_1 (r_{0L} J^{-1}_0 L^2) \nonumber \\
-  144 u_0^2 M^2_0 J_0^{-2}  L^{4-d} I_2 (r_{0L} J^{-1}_0 L^2), 
\end{eqnarray}
\begin{equation}
u^{eff}_0 =   u_0 
\left[ 1 -  36 u_0  \int_{\bf{k}} \hat{J}_{\bf{k}}^{-2} 
+ 36 u_0   J^{-2}_0  I_2 
(r_{0L} J^{-1}_0 L^2) L^{4-d} \right]. 
\end{equation}

The corresponding results of the field-theoretic model 
for $d > 4$ are obtained from Eqs. (70) and (71) as
\begin{eqnarray}
r^{eff}_0 &=&  (r_0 - r_{0c}) \left[ 1  -  12 u_0 \int_{\bf{k}} 
{\bf{k}}^{-4}\right]- 12 u_0  \; L^{2-d} I_1 (r_{0L}  L^2)  \nonumber \\
&-& 144 u_0^2 M^2_0  L^{4-d}  I_2 (r_{0L} L^2) - 12 u_0 \Lambda^{d-4} 
a_1 (d, r_{0L} \Lambda^{-2}) L^{-2}  \nonumber \\
&-& 144 u_0^2 \Lambda^{d-4}  M^2_0  a_2 (d, r_{0L} \Lambda^{-2})
(\Lambda L)^{-2}, 
\end{eqnarray}
\begin{eqnarray}
u^{eff}_0 \;  = \;  u_0 
\left [ 1 \;  - \; 36 u_0  \int_{\bf{k}} {\bf{k}}^{-4} 
\;  + \;  36 u_0 L^{4-d}  \; I_2 
(r_{0L} L^2) \right . \nonumber \\
\left . +36 u_0 \Lambda^{d-4} a_2 (d,r_{0L}\Lambda^{-2})(\Lambda L)^{-2}
\right ].
\end{eqnarray}
Substituting Eqs. (72) $-$ (75) into Eqs. (47) - (49) completes our 
calculation of the asymptotic form of $H^{eff}$ 
and of the order-parameter distribution function $P(\Phi)$, Eq. (45), 
in one-loop order for $d > 4$ and $n = 1$. The restriction "asymptotic"
means that these results, Eqs. (72)-(75), are applicable to arbitrary 
$r_{0L}L^2$
only in the sense that $L/\tilde{a}$ must be large and $|t|$ must be small.

As found already in the  large-$n$ limit \cite{CD1},
the leading ``shift of $T_c$'' [3] in $r^{eff}_0$, Eq. (72),  
is not just a temperature independent constant
$\sim L^{2-d}$ for the lattice model but a more complicated function
of $r_{0L} L^2$. For the field-theoretic model 
the leading shift is proportional to
$L^{-2}$ according to Eq. (74) 
and is  nonuniversal, i.e., explicitly cutoff-dependent. 
This result differs from the (temperature independent) shift $\sim L^{2-d}$
predicted for the field-theoretic model \cite{B1} and from the shift 
$\sim L^{-d/2}$ considered in previous work [7$-$9].
Our shifts are
caused by the (inhomogeneous) higher modes of the order-parameter fluctuations.
They cannot be neglected even for large $L/\tilde{a}$ (except for the extreme 
case of the bulk limit) and cannot be regarded only as ''corrections'' to the 
lowest-mode approximation  since they lead to a two-variable finite-size 
scaling structure for both the  field-theoretic and the lattice model, 
in contrast
to the single-variable scaling structure of the lowest-mode approximation,
as will be further discussed in Section 4.2.

Our results can be generalized to $n > 1$ by means of a nonperturbative
treatment of the order-parameter distribution function of the 
$O(n)$ symmetric $\varphi^4$ theory \cite{CDS}. It is obvious that this
does not change the conclusions regarding the structural differences between
the finite-size effects of the field-theoretic and lattice versions of the
$\varphi^4$ model.

\subsection*{4. Finite-size scaling functions of the lattice model}

\subsubsection*{4.1   Analytic results}

In the following we consider only the case $h = 0$.
Inspection of the asymptotic expressions of $r^{eff}_0$ and $u^{eff}_0$,
Eqs. (72) and (73), shows that they depend on three different lengths
$\xi_0$, Eq.(29), $l_0$, Eq.(30),  and $L$.
Therefore there exist different ways of writing $H^{eff}$ in a finite-size
scaling form.

Considering the ratio $\xi/L$ as a fundamental dimensionless variable [13-15]
leads to the following scaling form
\begin{equation}
H^{eff} \; = \; F(z,x,y)
\; = \; \frac{1}{2} r^{eff} (x,y) z^2 \; + \; u^{eff} (x,y) z^4 \; ,
\end{equation}
where $x$ and $y$ are the scaling variables given in Eqs. (1) and (2) and $z$
is the scaled order-parameter variable
\begin{equation}
z = \; J^{-1/2}_0 \; L^{(d-2)/2} \Phi \quad .
\end{equation}
The scaled effective parameters are in one-loop order 
\begin{eqnarray}
r^{eff} (x,y) \; = \; r^{eff}_0 L^2 J^{-1}_0 \; = \;
x - 12 I_1 (\bar{r})y - 144 \vartheta_2 (y_0) I_2 (\bar{r})y^{3/2}
\end{eqnarray}
and
\begin{eqnarray}
u^{eff}(x,y)\; = \;  u^{eff}_0 \; L^{4-d} J^{-2}_0 \; = \; y + 36 I_2
(\bar{r})y^2  
\end{eqnarray}
with
\begin{equation}
\bar{r} \; = \; x + 12 \vartheta_2 (y_0)y^{1/2} \;  
\end{equation}
and
\begin{equation}
y_0 \; = \; xy^{-1/2} \quad .
\end{equation}
This  leads to the finite-size scaling form
\begin{equation} 
\chi^{\pm} = L^2 P^{\pm}_\chi (x,y)
\end{equation}
and
\begin{equation}
M = L^{(2-d)/2} P_M (x,y) 
\end{equation}
with the two-variable scaling functions
\begin{eqnarray}
P^+_\chi (x,y) &=&  J^{-1}_0 \left[u^{eff}(x,y)\right]^{-1/2}
\vartheta_2 (Y(x,y)), \\
P^-_\chi (x,y) &=&  J^{-1}_0 \left[u^{eff}(x,y)\right]^{-1/2}
\left\{ \vartheta_2 (Y(x,y)) - \left[ \vartheta_1 (Y(x,y))\right]^2 
\right\}, \\
P_M (x,y) &=&  J^{-1/2}_0  \left[ u^{eff}(x,y)\right] ^{-1/4}
\vartheta_1 (Y(x,y)), \\
U(x,y) &=&  1 - \frac{1}{3}  \frac{\vartheta_4 (Y(x,y))}
{\left[\vartheta_2(Y(x,y))\right]^2},
\end{eqnarray}
where 
\begin{equation}
Y(x,y) \; = \; r^{eff} (x,y) \; 
\left[u^{eff}(x,y)\right]^{-1/2} \quad .
\end{equation}
The traditional finite-size scaling theories for $d < 4$ [13-18] have no 
asymptotic dependence on a second scaling variable $y$. We see that our 
scaling functions do not depend on the nonuniversal 
model parameters $\tilde{a}, J_{ij}, a_0, u_0$, except via the length scales
$\xi_0$ and $l_0$ contained in $x$ and $y$, and apart from the metric 
prefactors $J_0^{-1}$ and $J_0^{-1/2}$ in Eqs. (84)-(86).
Thus we may consider these functions to be universal in a restricted
sense, i.e., for a certain class of lattice models (rather than continuum
models, see below). 

From the previous one-loop finite-size theory \cite{EDC} and the successful
comparison with high-precision MC data in three dimensions \cite{CDT} it has 
become clear that careful consideration must be devoted to the appropriate 
form 
of evaluating these one-loop results. The previous analysis indicated that 
the prefactor $(u^{eff})^{-1/2}$ in Eqs. (84)-(88) should be further expanded 
with respect to the coupling $u_0$, in the spirit of a systematic perturbation 
approach, see Eqs. (5.45), (5.46) and footnote $1$ of Sec.$7$ of Ref. 
\cite{EDC}, as well as Eqs. (6.15), (6.16), (6.31) and (6.32) of Ref. 
\cite{EDC}. Thus the expanded forms $(u^{eff})^{-1/2}=y^{-1/2}[1+18I_2(\bar{r})
y]^{-1}$ or $(u^{eff})^{-1/2}=y^{-1/2}[1-18I_2(\bar{r})y]$ should be 
substituted into Eqs. (84), (85) and (88), respectively [and similarly for
$(u^{eff})^{-1/4}$ in Eq. (86)]. These expanded forms should be taken into
account in a future quantitative comparison of Eqs. (84)-(88) with MC data.
In the present paper we confine ourselves, for simplicity, to the unexpanded
form of Eqs. (84)-(88), as has been done in the result for $U(x,y)$ presented
in Ref. \cite{CD3}. The same comment applies to Eqs. (99)-(103) below.

At $T_c \; (x=0)$ we obtain from Eqs. (78) - (88) the
large -$L$ behavior in one-loop order for $d > 4$ 
\begin{eqnarray}
\chi_c^+ & \;  \sim \; & L^2  J^{-1}_0 
\; y^{-1/2} \vartheta_2(0) \sim  L^{d/2} \quad , \\[1cm]
M_c & \; \sim \; & L^{(2-d)/2} J^{-1/2}_0 \; y^{-1/4} \vartheta_1 (0) \sim  
L^{-d/4} \quad , \\[1cm]
\lim\limits_{L \rightarrow \infty}U(0,y) & \; = \; & 
1 \; - \; \frac{1}{3} \vartheta_4(0)/\vartheta_2(0)^2 
\; = \;  0.2705 \quad . 
\end{eqnarray}
The exponents in Eqs. (89) and (90) and the asymptotic value in Eq. (91)
are identical with those obtained in the lowest-mode approximation at $T_c$
\cite{B1}. The dangerous irrelevant character of $u_0\sim y$ is clearly
exhibited in Eqs. (89) and (90) in the form of $y^{-1/2}$ and $y^{-1/4}$.

Alternatively we may employ, instead of $x$ and $y$, the variables 
$w$ and $y$ where $w$ is given by Eq. (3). This implies the 
following scaling form
\begin{equation}
H^{eff} \; = \; \tilde{F}(s,w,y) \; = \; \frac{1}{2} \tilde{r}^{eff}(w,y)
s^2 \; + \; \tilde{u}^{eff} (w,y) s^4
\end{equation}
with the scaled order-parameter variable
\begin{equation}
s\; = \; J^{-1/2}_0 \; L^{d/4} \; l_0^{(d-4)/4} \;  \Phi \quad .  
\end{equation}
The scaled effective parameters are
\begin{eqnarray}
\tilde{r}^{eff} (w,y) & =& r^{eff} y^{-1/2} \nonumber \\ 
&=& w - 12 I_1 (\tilde{r}) y^{1/2} 
- 144 \vartheta_2 (w) I_2 (\tilde{r}) y  
\end{eqnarray}
and
\begin{eqnarray}
\tilde{u}^{eff}(w,y) \; = \; u^{eff} y^{-1} \; = \; 1 + 36 I_2 (\tilde{r}) y
\end{eqnarray}
with
\begin{eqnarray}
\tilde{r} \; = \; [w + 12 \vartheta_2 (w)]y^{1/2} \quad .
\end{eqnarray}
This leads to the finite-size scaling forms
\begin{equation} 
\chi^{\pm} \; = \; L^{d/2} \tilde{P}^\pm_\chi (w,y)
\end{equation}
and 
\begin{equation}
M = L^{-d/4} \tilde{P}_M (w,y)
\end{equation}
with the two-variable scaling functions
\begin{eqnarray}
\tilde{P}^+_\chi(w,y) & = &   A \left[ \tilde{u}^{eff}
(w,y) \right]^{-1/2}\vartheta_2 (\tilde{Y} (w,y)), \\
\tilde{P}^-_\chi(w,y) & = &  
A \left[ \tilde{u}^{eff}(w,y) \right]^{-1/2} \left\{
\vartheta_2 (\tilde{Y} (w,y)) - [ \vartheta_1 (\tilde{Y} (w,y))]^2 
\right\}, \\
\tilde{P}_M(w,y) & = & 
 \sqrt{A} \left[ \tilde{u}^{eff}(w,y) \right]^{-1/4}
\vartheta_1 (\tilde{Y} (w,y))\\
\tilde{U}(w,y) & = & 
1  -  \frac{1}{3} \vartheta_4 (\tilde{Y}(w,y))/
\left[\vartheta_2 (\tilde{Y}(w,y)) \right]^2, 
\end{eqnarray}
where $A=J^{-1}_0 l_0^{(4-d)/2}$ and
\begin{eqnarray}
\tilde{Y}(w,y) \;& = & \;\tilde{r}^{eff} (w,y)
\left[ \tilde{u}^{eff}(w,y) \right]^{-1/2} \quad . 
\end{eqnarray}
In the lowest-mode approximation the $y$ dependence in Eqs. (92) - (103)
is dropped and Eqs. (94) and (103) are replaced by 
$\tilde{Y} = \tilde{r}^{eff} = u_0^{-1/2} a_0 tL^{d/2}$.

These results will be discussed and applied to $d = 5$
in the following Subsections.

\subsubsection*{4.2 Discussion}

Both sets of scaling variables $(x,y)$ and $(w,y)$ are useful in the analysis
of the finite-size scaling structure. First we consider the $(x,y)$ 
representation. In order to elucidate the effect of the fluctuations of
the (inhomogeneous) higher modes above and below $T_c$ we assume $y$ to be
small (large $L/l_0$) and  expand $\chi^+$
and $\chi^-$ with respect to $y$ at finite $|x| > 0$, i.e., $T \neq T_c$.
This yields
\begin{eqnarray}
\chi^+  &=&  \chi^+_b \; \left\{  1 \; - 12 \left[ x^{-1} \; - \;
I_1(x) \right] \frac{y}{x} \; + \;  O(y^2/x^2) \right\},\phantom{123} x > 0 \\ 
\chi^-  &=&  \chi^-_b \; \Big\{  1 \; + \; \left[
15 x^{-1} + 12 I_1 (-2x) \; - \; 36 x I_2 (-2x) \right]
 \frac{y}{x}   \nonumber \\ 
&&\phantom{12345}\; + \; O(y^2/x^2)  \Big\} , \phantom{123}   x < 0   
\end{eqnarray}
where $\chi^+_b$ and $\chi^-_b$ are the bulk quantities given in Eqs. (18)
and (21). Similar expressions can be derived for $M$ and $U$.

The terms $\sim x^{-1}$ in the square brackets can be traced back to the
lowest-mode contributions whereas the terms $\sim I_1(x)$ and $\sim I_m(-2x)$
arise from the higher (inhomogeneous) modes. If the latter terms were ignored
one could rewrite $\chi^\pm$ in a lowest-mode form with the single variable 
$x/y^{1/2}$, as noted previously in the case $n \rightarrow \infty$ [4$-$6].
For any finite $|x| = L^2/\xi^2$, however, i.e., along the straight dashed 
lines in the $L^{-1} - \xi^{-1}$ plane (Fig. 1), there exists no argument that
would allow one to ignore the $I_m$ terms arising from the higher modes. This
proves the necessity of including two separate scaling variables in general. 
In particular it is misleading to consider the finite-size effects of the 
higher
modes as a ''correction'' to the lowest-mode approximation $-$ in the same 
sense as changes of mean-field exponents caused by critical fluctuations
for $d < 4$ should not be considered as "corrections". The crucial
point is that the higher modes cause a new {\it{structure}} of the 
finite-size scaling functions that cannot be written in terms of a single
variable $x/y^{1/2}$ (except for the special case $x \rightarrow 0$,
$y \rightarrow 0$ at finite $x/y^{1/2}$, see below).
This structural aspect is a matter of principle, regardless of how large or 
small the effect of the higher modes might be.

In this context we take up a nontrivial aspect in the discussion in the
previous literature about role played by the ''shift of $T_c$''. It was
asserted that a term of the type $\sim L^{2-d}$ in the parameter 
$r^{eff}_0$
of $H^{eff}$ \cite{B1} represents a ''correction to scaling'' [9] or 
a "subdominant term" [12] that can be neglected in 
the large -$L$ limit compared to the lowest mode part $r_0 - r_{0c} = a_0 t$.
This assertion is incorrect, however, for the reasons just given in the
preceding paragraph. The term $\sim L^{2-d}$ in $r_0^{eff}$, Eq. (72), is,
in fact, the origin of the terms $\sim I_1(x)$ of $\chi^{\pm}$ in Eqs. (104)
and (105) which we have shown to represent nonnegligible contributions
rather than ''corrections''.
Similarly, the term $\sim M^2_0 L^{4-d}$ in $r^{eff}_0$, Eq. (72), is the
origin of the nonnegligible higher-mode contribution $I_2(-2x)$ to $\chi^-$
in Eq. (105).

The expansion in Eqs. (104) and (105) breaks down in the limit of small
$|x|$, i.e., large $\xi/L$. This includes the large -$L$ limit at 
$T = T_c$ where the exponent of the susceptibility $\chi_c \sim L^{d/2}$
and the Binder cumulant $U_c$ have been found [4$-$6] to agree with the
lowest-mode approximation for the lattice model (but not for the 
field-theoretic model). Here this result is seen from the representation in
terms of $w$ and $y$ as given
in Eqs. (92)-(103). In this representation the single-variable lowest-mode 
like structure 
appears as the leading $w$ dependence whereas the higher-mode contributions
$\sim I_1$ and $\sim I_2$ are multiplied by $y^{1/2}$ and $y$. 
As noted previously \cite{CD1} these higher-mode contributions are not
of a dangerous irrelevant character even though the dangerous irrelevant 
four-point coupling
$u_0$ determines the length scale 
$l_0 \sim u_0^{1/(d-4)}$. 

We shall see below that the $y^{1/2}$  terms are quantitatively important.
Nevertheless, at first sight it seems justified to consider the latter 
contributions
as asymptotically negligible in the limit $y \rightarrow 0$ corresponding to
$L \rightarrow \infty$. 
In the terminology of the renormalization group, this limit corresponds to
approaching the "Gaussian fixed point" of the dimesionless four-point
coupling $u_0 L^{4-d}$. Neglecting the $y$-dependence in this limit is 
justified, however, only if $w$
is kept finite, i.e., if $|t|$ vanishes sufficiently strongly. 
Keeping $w$ finite for $L \rightarrow \infty$ 
is a special case corresponding to  paths in the $L^{-1} - \xi^{-1}$ plane 
where the ratio $\xi/L$ diverges as
$L^{(d-4)/4}$. Such paths become asymptotically parallel to the
vertical axis (Fig. 1). This includes the special case $T = T_c$,
$L \rightarrow \infty$. The description of the entire 
$L^{-1} - \xi^{-1}$ plane, on the other
hand, requires both $w$ and $y$ as generic scaling variables in order to
correctly include the $w \rightarrow \infty$ limit which corresponds to 
a finite ratio $\xi/L$ for $L \rightarrow \infty$.
The latter limit $w \rightarrow \infty$ cannot be taken  correctly within 
the single-variable 
scaling structure such as $\tilde{P}_{\chi}^{\pm}(w,0)$ and within the reduced
scaling form $\tilde{F}(s,w,0)$.
This structure does not capture the complete finite-size effects
presented in Eq. (104) and (105) above. For this reason the inhomogeneous 
fluctuations must be considered as {\it relevant}, for finite $L$, in the 
sense of the renormalization group. For the same reason the reduced scaling 
form [equivalent to $\tilde{F}(s,w,0)$] that was proposed by Binder et al.
\cite{BNP} for general $\xi/L$ (not only for $\xi/L = \infty$), is not valid.

Our order-parameter distribution function $P(\Phi) \sim \exp (-H^{eff})$
can be compared with the zero-field probability distribution function of Binder
\cite{B2,14} below $T_c$
\begin{eqnarray}
P_L(s) \; = \; const 
\left \{ \; \exp \left[ - (s - m_b)^2 L^d/2\chi_b \right] \right. 
\nonumber\\[0.5cm]
\left. \; + \; \exp \left[ - ( s + m_b)^2 L^d /2\chi_b \right] \;\right \}
\end{eqnarray}
where $m_b \sim |t|^{1/2}$ and $\chi_b \sim |t|^{-1}$
are $L$ independent bulk quantities.
In $P_L(s)$ the temperature dependence enters in the
form $[L/l(t)]^d$ with the ''thermodynamic length'' [8,9,19]
$l(t) \sim |t|^{-2/d}$. 
Our theory identifies the relevant length scale $\tilde{\ell}$
of the corresponding variable $w$, Eq. (3), in terms of a combination of
$l_0$ and $\xi_0$ according to Eq. (4). 
The distribution function $P_L(s)$ has been invoked as an argument 
in support of the single-variable
scaling structure of the free energy (at $h = 0$) proposed by Binder
et al. \cite{BNP}. 
For finite $\xi/L$, however, our results do not agree
with the structure of  $P_L(s)$ which does not contain the important $y$ 
dependence reflected in the shift 
$\sim I_1 L^{2-d}$ in Eqs. (72), (78), (94) that is caused by the 
inhomogeneous modes. 
Thus the double Gaussian form of $P_L(s)$, Eq. (106), as well as the 
underlying theory of Gaussian thermodynamic 
fluctuations \cite{LL}, are not applicable to finite systems with periodic 
boundary conditions in the critical region above the upper critical 
dimension. This is remarkable in view of the fact that mean-field theory 
becomes exact in the bulk limit for $d > 4$.

\subsubsection*{4.3 Predictions for ${\bf d = 5}$}

We illustrate and further discuss our results for the lattice
model for the case $d = 5$
which can be compared with MC data of the five-dimensional Ising model.
In Fig. 2 we plot the order-parameter distribution function in terms of
$F$, Eqs. (76)-(81),
\begin{equation}
P(\Phi, t,L,u_0) \;  d \Phi \; = \; 
\frac{\exp [-F(z,x,y)]}{\int\limits_{-\infty}^\infty dz \exp 
\left[ - F (z,x,y)\right]} \;  dz \; ,
\end{equation}
for typical values of $x$ and $y$ above, at and below $T_c$. 
The shape of these functions resembles that of the MC data shown
in Fig. 1 of Ref. \cite{RNB}. In order to demonstrate the effect of the 
higher modes on these functions we show the order parameter distribution 
function in Fig. 3 in terms of $\tilde{F}$, Eqs. (92)-(96),
\begin{equation}
P(\Phi, t,L,u_0) \;  d \Phi \; = \; 
\frac{\exp [-\tilde{F} (s,w,y)]}{\int\limits_{-\infty}^\infty ds \exp 
\left[ - \tilde{F} (\tilde{s},w,y)\right]} \;  ds \; ,
\end{equation}
with given $w$ but for several values of $L/l_0$ including the limiting
function for $L/l_0 \rightarrow \infty$ at fixed $w$. The latter function
has the structure $P_0 (\Phi)$  of the lowest-mode
approximation (with one-loop expressions for the reference lengths $\xi_0$ and
$l_0$). This does not mean, however, that $P_0(\Phi)$ is the exact
representation of $P$ in the large -$L$ limit in general.
The constraint $w = const < \infty$ restricts the validity of $P_0(\Phi)$
only to the special region $|t|L^{d/2} < \infty$ corresponding to
$\xi/L \rightarrow \infty$ (region between the curved dotted lines in Fig. 1).
The width of this region in the $L^{-1}-\xi^{-1}$ plane vanishes 
asymptotically for $L \rightarrow \infty$. This special region is of interest 
because it contains characteristic
("pseudocritical" [13]) temperatures close to $T_c$ such as $T_{max}(L)$ 
and $T_c(L)$ to be defined below.

The scaling functions $\tilde{P}_{\chi}^{\pm}$, Eqs. (99) and (100), 
of $\chi^+$ and $\chi^-$ corresponding to the 
order-parameter distribution function of Fig. 3  are
shown in Fig. 4. Similar plots can be made for $M$ and $U$. For comparison 
with MC data we refer to Figs. 11-14 of Ref. [9]. At first sight, 
the changes due to the variation of
$L/l_0$ appear to be small.
At the level of accuracy of previous MC data \cite{RNB}, however,
these changes and their disagreements with the lowest-mode predictions
\cite{B1} have been clearly detected and have been considered as a major
discrepancy, regardless of their smallness, because of their unexplained 
weak $L$ dependence.

Here we further elucidate the deviations from the lowest-mode
predictions by plotting in Fig. 5 the scaling functions of 
$U, \chi^\pm$ and $M$ at and below 
$T_c$ as functions of the
reduced length $L/l_0$. As originally found in Ref. \cite{CD3} for the
example of the Binder cumulant at $T_c$, the slow 
approach to the asymptotic $(L \rightarrow \infty)$ values arises from the
$y^{1/2} \sim L^{(4-d)/2}$ terms. This slow approach was observed in the 
previous MC data \cite{RNB} and, at that time, gave sufficient reason to doubt
the correctness of the lowest-mode predictions \cite{B1} which do not have 
a weak subleading $L$-dependence at $T_c$. Our theory now shows that 
subsequent attempts [10-12] to explain the discrepancies did not resolve the
problems.  
In particular, the bulk form of the renormalization-group flow equations 
employed by Bl\"ote and Luijten \cite{LB}
led to an apparent confirmation of the (incorrect) shift 
$\sim L^{2-d}$ predicted in Ref. \cite{B1} for the {\it field-theoretic} 
model. These bulk flow equations \cite{LB} do not correctly describe the 
finite-size effects of the $\varphi^4$ {\it lattice} model either.
We believe that our theory identifies the origin of the
previous discrepancy, apart from possible quantitative aspects which we shall
address elsewhere, after a quantitative identification of the lengths 
$\xi_0$ and $l_0$.

An interesting consequence of the existence of the limiting function 
$P_0 (\Phi)$ 
mentioned above is the existence of limiting scaling functions such as 
$\tilde{P}_\chi^- (w,0)$ of $\chi^-$ for $L/l_0 \rightarrow \infty$ 
at fixed $w$ [Fig. 4 (b)]. For finite $L$, $\chi^-$  
exhibits a maximum $\chi_{max}$ below $T_c$ at a temperature $T_{max}(L)$. The
asymptotic $L$ dependence of $T_c - T_{max}(L)$ can be inferred from the 
fact that $\tilde{P}_\chi^- (w,0)$ has a temperature dependence only of the 
form of $w$ $\sim t L^{d/2}$.
This implies the large -$L$ behavior $T_c - T_{max}(L) \sim L^{-d/2}$
and correspondingly $\chi_{max}\sim L^{d/2}$.
Similar arguments lead to our prediction $T_c - T_c(L) \sim L^{-d/2}$
where $T_c(L)$ is the ''effective critical temperature'' \cite{RNB}
at which the magnetization has its maximum slope. The same power law is valid
for the temperature at which the specific heat has its maximum. These power 
laws $\sim L^{-d/2}$  agree with the MC data \cite{RNB}. The true asymptotic
amplitudes of these power laws, however, have not been observed in previous 
MC simulations in $d = 5$ dimensions because of the slow approach of the 
subleading terms $\sim L^{(4-d)/2}$ towards $L \rightarrow \infty$ mentioned 
above. MC simulations of larger systems would be desirable for testing the
magnitude of such subleading terms predicted by our theory.

As indicated already in Figs. 3 and 4 of Ref. [6], we also point to an
additional interesting effect of practical importance. In Fig. 6 we have
plotted the Binder cumulant as a function of $x$  
for several values of $L/l_0$. Without the effect of the slowly decaying
contribution $\sim y ^{1/2} \sim L^{(4-d)/2}$ one would have 
expected \cite{RNB}
a well identifiable intersection point  of these curves if $L$ is, 
say, larger than
$10\; \tilde{a}$. For $d < 4$, this features has been a standard and successful
empirical  method of determining the value of bulk $T_c$ from MC data of
finite systems. 
Our previous \cite{CD3} and present figures demonstrate that this method is not
accurately applicable, without additional information, to systems with $d = 5$.

\subsection*{5. Finite-size scaling functions of the ${\bf{\varphi^4}}$ field
theory}

For the field-theoretic model the effective parameters $r^{eff}_0$ and
$u^{eff}_0$, even in their asymptotic form in Eqs. (74) and (75), depend 
explicitly on the length 
$\Lambda^{-1}$, in addition to the lengths $\xi_0, l_0$ and $L$, as  
found already in the limit $n \rightarrow \infty$ \cite{CD1}. 
This means that none of the original nonuniversal model parameters
$a_0, u_0$ and $\Lambda$ becomes unimportant even close to $T_c$ and for
large $L$. In a scaled form the effective parameters read
\begin{eqnarray}
r^{eff} = r^{eff}_0 \; L^2 \; & = & \;x - 12 \;  I_1 (\bar{r})y \; - 144 \; 
\vartheta_2
(y_0) I_2 (\bar{r}) y^{3/2}\nonumber \\
& - & 12 u_0 \Lambda^{d-4} \; a_1 
(d, \bar{r}   \Lambda^{-2} \;)  \nonumber \\
& - & 144 (u_0 \Lambda^{d-4})^{3/2} \vartheta_2 (y_0) a_2 
(d, \bar{r} \Lambda^{-2} ) (\Lambda L)^{-d/2} \;  
\end{eqnarray} 
and 
\begin{eqnarray}
u^{eff} \;& = &\; u_0^{eff} \; L^{4-d} \; = y 
+ 36 I_2 (\bar{r})y^2\nonumber \\[1cm] 
& + & \; 36 (u_0 \Lambda^{d-4})^2 \; a_2 (d, \bar{r} \Lambda^{-2})
(\Lambda L)^{2-d} 
\end{eqnarray}
where $\bar{r}$ and $y_0$ are given in Eqs. (80) and (81).
The last term $\sim L^{2-d}$ in Eq. (110) can be neglected asymptotically.

Substituting these expressions into Eqs. (76), (84) $-$ (88) and (107)
(with $J_0 = 1$) yields the finite-size
scaling functions of the order-parameter distribution function and of 
the quantities $\chi^{\pm}, M$ and $U$.
The two-variable finite-size scaling functions depend on $x$ and $y$ and,
in addition, explicitly on the nonuniversal parameter $u_0 \Lambda^{d-4}$.
Thus the scaling functions are nonuniversal for $n = 1$, and obviously
also for general $n$.
At $T_c$, the asymptotic power laws of $\chi$ and $M$ are found to be
\begin{eqnarray}
\chi^+_c (L) \; &=& \; L^2 P^+_\chi (0,y) \sim L^{d-2}\quad ,\\[1cm]
M_c(L) \; &=& \;L^{(2-d)/2}\; P_M (0,y) \sim L^{-1} 
\end{eqnarray}
for the field-theoretic model which differ from those of the lattice
model in Eqs. (89) and (90). The asymptotic value of $U$ at
$T_c$ for the field-theoretic model is in one-loop order
\begin{equation}
\lim\limits_{L \rightarrow \infty} \; U(0,y) \; = \; 2/3
\end{equation}
which is far from that of the lattice model in Eq. (91).
The significant differences between Eqs. (89)-(91) and Eqs. (111)-(113) 
are due to the $L$-independent but cutoff-dependent term 
$\sim u_0 \Lambda^{d-4}$ in Eq. (109), similar to the constant additive term
in Eq. (122) of Ref. \cite{CD1}.
Existing MC data for Ising models on $d = 5$ lattices (such as 
the MC result $U_{MC}=0.319\pm 0.017$ in Ref. \cite{RNB}) clearly disagree 
with these 
field-theoretic results and  rule out the possibility that the
$\varphi^4$ field theory provides a correct description of finite lattice
systems above the upper critical dimension. 
In particular, the prediction of a breakdown of
universality for finite systems above the upper critical dimension 
constitutes a serious failure of the continuum approximation for lattice
systems. The present results for finite $n$ confirm our earlier 
[4-6,22] assertion regarding the applicability of the $\varphi^4$ field theory 
for $d > 4$.

\subsection*{6. Summary and conclusions}

We summarize and further comment on the results of this paper as follows.

On the basis of a one-loop calculation for the $\varphi^4$ model on a lattice
and for the $\varphi^4$ continuum model
in a cubic geometry with periodic boundary conditions above four dimensions
we have shown that our general conclusions regarding universality and
finite-size scaling  inferred from the large-$n$ limit
[4-6] remain valid for finite $n$.  In particular, $\varphi^4$ field theory
based on the Landau-Ginzburg-Wilson continuum Hamiltonian \cite{ZJ} 
does not correctly
describe the leading finite-size effects of spin systems on a lattice
with $d > 4$.

Although the critical exponents of mean-field theory are exact for bulk systems
above four dimensions, the thermodynamic theory of Gaussian  fluctuations
\cite{LL} is not applicable to finite systems with periodic boundary conditions
in the critical region for $d > 4$.

Finite-size scaling in terms of a single scaling variable, as predicted by the
phenomenological theory of Binder et al. \cite{BNP} and by the lowest-mode
approximation of Br\'{e}zin and Zinn-Justin \cite{B1}, is not valid for the
$\varphi^4$ field theory. For the $\varphi^4$ lattice model it is not valid
for any finite $\xi/L$ where $\xi$ is the bulk correlation length. As 
originally conjectured in Ref. \cite{15}, lowest-mode like large -$L$ behavior
is asymptotically correct for the lattice model at $T_c$ as shown previously
for the susceptibility [4$-$6] and the Binder cumulant \cite{CD3}; furthermore,
it is valid in the small region of finite $|w| \sim |t| L^{d/2}$
in the large -$L$ and small $|t|$ limit (Fig. 1). This region,  
corresponding to a divergent
ratio $\xi/L \sim L^{(d-4)/4} \rightarrow \infty$, represents only a 
small part (between the two curved dotted lines of Fig.1) of the general 
finite-size scaling regime (of arbitrary finite $\xi/L$)  for which earlier
theories \cite{B1,BNP} were originally thought to be valid. Our two-variable
finite-size scaling structure is consistent with that proposed by Privman and
Fisher \cite{PF} but is significantly less universal than anticipated 
previously [15].

The inhomogeneous higher modes have been shown to be relevant above the
upper critical dimension, contrary to different statements in the previous
literature [1,3,10,12,15,27-43]. We have identified the characteristic length 
scale $l_0$, Eq. (30), of the finite-size effects of the higher modes in 
terms of
the amplitude of the bulk correlation length $\xi$ at $T = T_c$ for small
external field $h$. The one-loop finite-size effects 
arising from the relevant
higher modes do not represent ''corrections'' to the lowest-mode 
approximation but constitute a generic part of the correct finite-size 
scaling structure.
By contrast, two-loop contributions  are expected to represent only 
quantitative corrections that  will not change the scaling structure.

The ''shift of $T_c$ '' \cite{B1}  in the temperature variable $r_0^{eff}$ 
of the exponential
order-parameter distribution function is proportional to $L^{-2}$ for the
field-theoretical model and proportional to $I(t,L^{-1})L^{2-d}$ for
the lattice model where the function $I(t,L^{-1})$ has a finite limit
$I(0,0)$. The effects caused by these  shifts remain nonnegligible at
any finite ratio $\xi/L$ even in the large -$L$ limit as demonstrated 
in Eqs. (104) and (105) for the example of the susceptibility above and
below $T_c$.

The ''shift of $T_c$'' $\sim L^{2-d}$ mentioned in the preceding paragraph 
must be distinguished from shifts of 
characteristic temperatures, in the sense of pseudocritical temperatures
\cite{F}, such as the temperature $T_{max}(L)$
at which the susceptibility $\chi^-(t,L)$ has its maximum, or the
''effective critical temperature'' $T_c(L)$ \cite{RNB} where the 
order parameter has its maximum slope. We find that these ''shifts'' have the
asymptotic (large $L$) behavior $T_c - T_{max}(L) \sim L^{-d/2}$
and $T_c - T_c(L) \sim L^{-d/2}$ for the $\varphi^4$ lattice model. Similarly
our theory implies $\chi_{max}\sim L^{d/2}$ asymptotically.
This is a simple consequence of the fact that the order-parameter distribution
function shown in Fig.3 has a finite limit for $L \rightarrow \infty$ at 
finite $w$ and that the position of $T_{max}(L)$
and $T_c(L)$ remain located in the temperature region of finite $w$ in the 
limit $L \rightarrow \infty$.

Our theory identifies the possible origin of a significant discrepancy 
between MC
data at $d = 5$ [9] and the lowest-mode prediction for the Binder cumulant at
$T_c$ \cite{B1} in terms of slowly
decaying finite-size terms $\sim L^{(4-d)/2}$ (Fig. 5).  
These terms also mask
the true asymptotic amplitudes of the power laws $\chi_c \sim L^{d/2}$ and
$M_c \sim L^{-d/4}$ at $T_c$.
For the same reason
the method of determining bulk $T_c$ (from MC data via the intersection point
of the Binder cumulant) is demonstrated in Fig. 6 to become 
quantitatively inaccurate at $d = 5$, as found originally in Ref. \cite{CD3}.

Quantitative predictions for various asymptotic finite-size scaling 
functions have been made for $d = 5$ and $n = 1$ (Figs. 2$-$6). 
These predictions are expected to be  valid for sufficiently
large $L/\tilde{a}$ and small $|t|$. The true range of applicability
remains to be explored by quantitative comparisons  with MC data, after an
appropriate identification of the nonuniversal lengths $\xi_0$ and $l_0$.
As noted previously \cite{CD1}, it is not yet established whether the 
$\varphi^4$ model
on a finite lattice is fully equivalent to finite  spin models  
regarding the leading
and subleading finite-size effects.

Our results indicate that in the limit $d-4 \rightarrow 0^+$, for
systems with periodic boundary conditions, different amplitudes of 
finite-size effects at $d=4$ are obtained depending on whether a lattice
model or a continuum model is considered. In view of this possible 
ambiguity at $d=4$ the limiting behaviour for $4-d \rightarrow 0^+$ (i.e.,
$\epsilon \rightarrow 0^+$ in the standard $\epsilon=4-d$ expansion ) should
also be reexamined for lattice models at finite lattice spacing and for
continuum models at finite cutoff.

\vspace{5mm}

{\bf{Note added}}

After completion of the present work we received a preprint 
'' Finite-size scaling above the upper critical dimension revisited:
The case of the five-dimensional Ising model'' by E. Luijten, 
K. Binder, and H.W.J. Bl\"ote where 
the authors compare the asymptotic result for $U(x,y)$ in the (unexpanded)
form of Eqs. (87) and (88) with
their  Monte Carlo data of the five-dimensional  Ising model.
The authors confirm the '' occurrence of spurious cumulant intersections''
predicted in Figs. 3 and 4 of Ref. \cite{CD3} and agree with the slow 
convergence of finite-size effects
for $L \rightarrow \infty$ found in Ref. \cite{CD3} which essentially
resolves the longstanding discrepancies noted in the MC studies in 
Refs. [7-9] regarding the Binder cumulant. 
On a much more quantitative level than considered previously [3-12],
however, the authors estimate the length $l_0$ as $l_0 = 0.603 \; (13) $ and 
claim to find {\it{new}}  ''significant 
discrepancies'' between their MC data for small $L$ and our asymptotic 
(large $L$) one-loop result for $\chi^+$ in the (unexpanded) form of 
Eq. (84).

We doubt the significance of the quantitative deviations  for small
system sizes $L = 4$ and $8$ shown in their Figs. 7(a) and (b), except for the 
{\it{sign and curvature}} of the deviations
from the large-$L$ behavior of the susceptibility shown 
in their Figs. 8 and 9. We propose that the latter issue can be resolved 
essentially on the basis of our complete {\it non-asymptotic} one-loop 
expression for $H^{eff}$ presented in Eqs. (47)-(49) of this paper or on the 
basis of the underlying order-parameter distribution function \cite{CDS}, 
rather than by an asymptotic two-loop calculation suggested 
by Luijten et al. We note that similar non-asymptotic effects are well known
for small spin systems at $T_c$ in three dimensions as discussed in the 
context of Fig. 14 of Ref. [18]. 

We doubt the reliability of the estimate of $l_0=0.603 \; (13)$ by Luijten 
et al. since it was found by applying the asymptotic ($L \rightarrow \infty$) 
expression for $\chi_c^{+}$ in their Eq. (31) to non-asymptotic MC data. 
Our Fig. 5b indicates that the apparent (6 percent) mismatch between 
theory and MC data for $L \leq 22$ in Fig. 9 of Luijten et al. is due to this 
inadequate estimate of $l_0$.

Part of the remarks by Luijten et al. regarding the limiting case 
$t \rightarrow 0$, 
$L \rightarrow \infty$ at fixed $t L^{d/2}$ agree with our earlier and present 
independent 
findings. We disagree, however, with their claim that 
''there is no contradiction at all'' between our finite-size theory 
and the ideas of Ref. \cite{BNP}. 
First, we note that the ideas of Ref. [7] fail for the continuum $\varphi^4$ 
model. Second, we maintain that the single-variable scaling 
structure (for $h=0$) proposed in Ref. \cite{BNP}
does not capture the correct structure of finite-size effects of the lattice
model at any finite
value of $\xi/L$ (see Fig.1). In particular the ideas of Ref. [7] do not lead
to a correct description of the finite-size departures from bulk critical 
behavior at small but fixed $|t|$
in the asymptotic range $0 < |t| \ll 1$ [compare Eqs. (104) and (105)] to 
which an acceptable finite-size
scaling structure should be applicable (such as the general structure proposed 
in Ref. \cite{PF}).

\vspace{5mm}

{\bf{Acknowledgment}}

We thank  K. Binder, H.W.J. Bl\"ote, and E. Luijten for sending us their 
preprint.
Support by Sonderforschungsbereich 341 der Deutschen Forschungsgemeinschaft
and by NASA under contract number 960838 and 100G7E094 is acknowledged. 
One of the authors (X.S.C.) thanks the National Natural Science Foundation 
of China for support under Grant No.
19704005. We also thank the referee for useful comments.

\newpage
{\bf{Figure Captions}}

{\bf{Fig.1.}} Asymptotic $L^{-1} - \xi^{-1}$ plane (schematic plot) 
above $(t > 0)$ and below $(t < 0)$
$T_c$ for the lattice model above four dimensions where 
$L$ is the system size and $\xi$ is the bulk correlation
length (in units of the lattice spacing). The straight dashed lines 
correspond
to paths at constant finite ratio $\xi/L$. 
The curved dotted lines represent
paths at constant finite $t L^{d/2}$ for $L \rightarrow \infty$ and
$t \rightarrow 0$ corresponding to a divergent ratio
$\xi/L \sim 
L^{(d-4)/4} \rightarrow \infty$.
The single-variable scaling form in terms of $w = t (L/\tilde{l})^{d/2}$, 
Eqs. (3) and (4),
is valid only in the region between the curved dotted lines. The two-variable
scaling form in terms of 
$x = t L^2/\xi^2_0$ and $y = (L/l_0)^{4-d}$ or $w$ and $y$ is necessary 
in the entire $L^{-1} - \xi^{-1}$ plane where $\xi/L$ is finite.

{\bf{Fig. 2.}} Theoretical prediction of the scaled order-parameter 
distribution function $P(\Phi,t,L,u_0) L^{-(d-2)/2} J_0^{1/2}$ 
of the $\varphi^4$ lattice model in $d = 5$ dimensions ${\mbox{vs}} 
\; z$,
Eq. (77), in the form of Eq. (107) where $F(z,x,y)$ is given by Eqs. (76)$-$ 
(81). The parameter values are $x = 0.25$ above $T_c \; ({\bf{a}})$, 
$x = 0$ at $T_c \; ({\bf{b}})$,
and $x = - 0.75$ below $T_c \;  ({\bf{c}})$, where $|x| = L^2/\xi^2$.
The reduced length is  $y^{-1} \, = \; L/l_0 = 16$. Compare Fig. 1 of 
Ref. \cite{RNB}.

{\bf{Fig.3.}} 
Theoretical prediction of the scaled order-parameter distribution
function $P(\Phi,t,L,u_0)L^{-d/4} J^{1/2}_0 \; l_0^{(4-d)/4}$ of the 
$\varphi^4$ lattice model in $d = 5$ dimensions 
${\mbox{vs}} \;  s$,
Eq. (93), in the form of Eq. (108) where $\tilde{F}(s,w,y)$ is given by
Eqs. (92) $-$ (96). The parameter values are 
$w = 1$ above $T_c \; ({\bf{a}})$, $w = 0$ at $T_c \; ({\bf{b}})$, 
and $w = - 3$
below $T_c \; ({\bf{c}})$ where $w = t (L/\tilde{\ell})^{d/2}$.
The curves are shown for two reduced lengths $L/l_0 = 8$
(dotted lines) and $L/l_0 = 32$ (dashed lines) as well as for the limiting 
case $L/l_0 = \infty$ (solid lines)
at fixed $w$.

{\bf{Fig. 4.}} Theoretical prediction of the finite-size scaling functions
\newline
$\tilde{P}^+_\chi (w,y)J_0 l_0^{(d-4)/2}$ and 
$ \tilde{P}^-_\chi (w,y)J_0 l_0^{(d-4)/2}$  of the susceptibilies
$\chi^+$, Eq. (99), $({\bf{a}})$ and $\chi^-$, Eq. (100), $({\bf{b}})$
of the lattice model in $d = 5$ dimensions
${\mbox{vs}} \; w = t (L/\tilde{\ell})^{d/2}$ for several 
values of the reduced length
$L/l_0$, corresponding to Fig. 3, including the limiting case $L/l_0 = \infty$
at fixed $w$. In this representation the position of the maximum of 
$\tilde{P}^-_\chi$ attains a finite value for 
$L/l_0 \rightarrow \infty$  which determines $T_c -T_{max} \sim L^{-d/2}$.
Compare Fig. 13 of Ref. [9].

{\bf{Fig. 5.}}  Theoretical prediction of various finite-size scaling 
functions 
of the lattice model in $d = 5$ dimensions, 
$U(0,y)$ $({\bf{a}})$  , $\tilde{P}^+_\chi (0,y)J_0 l_0^{(d-4)/2}$ 
$({\bf{b}})$, 
$\tilde{P}_M(0,y)J_0^{1/2} l_0^{(d-4)/4}$
$({\bf{c}})$, $  \tilde{P}^-_\chi (-3,y)J_0 l_0^{(d-4)/2}$ $({\bf{d}})$ , 
$  \tilde{P}_\chi^- (-6,y)J_0 l_0^{(d-4)/2}$ $({\bf{e}})$ according to 
Eqs. (99)-(103), as a function of the scaled length 
$y^{-1} = L/l_0$. The  arrows indicate the asymptotic
one-loop values for $L \rightarrow \infty$. Compare Fig.2 of Ref.[6].

{\bf{Fig. 6.}}  Theoretical prediction of the Binder cumulant $U(x,y)$,
Eq. (87), as a function of the scaled reduced temperature $x = t(L/\xi_0)^2$,
Eq. (1), in the range $-4 \leq x \leq 4 \; ({\bf{a}})$ and in the range
$-0.4 \leq x \leq 0.4 \; ({\bf{b}})$, for several values of the 
scaled length $y^{-1} = (L/l_0)^{d-4}$, 
at $d = 5$ : 
$L/l_0 = 8$ (dotted line), 
$L/l_0 = 16$ (dot-dashed line), 
$L/l_0 = 32$ (dashed line).
The bulk value [thin solid line in $({\bf{a}})]$ is $2/3$ below $T_c$ and $0$
above $T_c$.
The cross indicates the asymptotic one-loop value $U(0,0) = 0.2705$ at
$T_c, L \rightarrow \infty$, Eq. (91). Compare Figs. 3 and 4 of 
Ref. \cite{CD3}.


\newpage
\begin{thebibliography}{99}
\bibitem{ZJ}
J. Zinn-Justin,{\it{ Quantum Field Theory and Critical Phenomena}} 
(Clarendon Press, Oxford, 1996).
\bibitem{BGZJ}
K.G. Wilson and J. Kogut, Phys. Rep. {\bf 12C}, 74 (1974);
M.E. Fisher, Rev. Mod. Phys. {\bf 46}, 597 (1974); {\bf 70}, 653 (1998);
E. Br\'{e}zin, J.C. Le Guillou, and J. Zinn-Justin, in {\it{Phase Transitions
and Critical Phenomena}}, Vol. 6, eds. C. Domb and M.S. Green (Academic Press,
New York, 1976), p. 125.
\bibitem{B1}
E. Br\'{e}zin and J. Zinn-Justin, Nucl. Phys. {\bf{B257}}[FS14], 867 (1985).
\bibitem{CD1}
X.S. Chen and V. Dohm, Eur. Phys. J. {\bf B 5} (1998), in press.
\bibitem{CD2}
X.S. Chen and V. Dohm, Physica {\bf{A 251}}, 439 (1998)
\bibitem{CD3}
X.S. Chen and V. Dohm, to be published in Int. J. Mod. Phys. {\bf C}.
This paper was presented at the Satellite Meeting to STATPHYS 20  
''{\it{Applications of Field Theory to Statistical Physics}}'', Bonn,
July 15-18, 1998.
\bibitem{BNP}
K. Binder, M. Nauenberg, V. Privman, and A.P.Young,
Phys. Rev. {\bf{B 31}}, 1498 (1985).
\bibitem{B2}
K. Binder, Z. Phys. {\bf{B 61}}, 3 (1985).
\bibitem{RNB}
Ch. Rickwardt, P. Nielaba, and K. Binder, 
Ann. Phys. (Leipzig)  {\bf{3}}, 483 (1994).
\bibitem{LB}
E. Luijten and H.W.J. Bl\"ote, Phys. Rev. Lett. {\bf{76}}, 1557 (1996);
H.W.J. Bl\"ote and E. Luijten, Europhys. Lett. {\bf{38}}, 565 (1997);
E. Luijten, Europhys. Lett. {\bf{37}}, 489 (1997).
\bibitem{M}
K.K. Mon, Europhys. Lett. {\bf{34}}, 399 (1996); ibid. {\bf{37}}, 493 (1997).
\bibitem{PRL}
G. Parisi and J.J. Ruiz-Lorenzo, Phys. Rev. {\bf{B54}}, R 3698 (1996); 
{\bf{B55}}, 6082 (1997).
\bibitem{F}
M.E. Fisher, in {\it{Critical Phenomena}}, International School of Physics
''Enrico Fermi'', Course 51, ed. M.S. Green (Academic, New York, 1971).
\bibitem{B}
M.N. Barber, in {\it{Phase Transitions and Critical Phenomena}}, Vol. 8, eds.
C.Domb and J.L. Lebowitz (Academic Press, New York, 1983), p. 145.
\bibitem{Pri}
V. Privman, in {\it{ Finite Size Scaling and Numerical Simulation of
Statistical Systems}}, ed. V. Privman (World Scientific, Singapore, 1990),
p.1.
\bibitem{Bi1}
K. Binder, Annu. Rev. Phys. Chem. {\bf{43}}, 33 (1992).
\bibitem{Bi2}
K. Binder, in {\it{ Computional Methods in Field Theory}}, eds. H. Gausterer
and C.B. Lang (Springer, Berlin, 1992), p. 59.
\bibitem{VD} V. Dohm, Physica Scripta {\bf T 49} , 46 (1993).
\bibitem{14}
K. Binder, Z. Phys. {\bf{B43}}, 119 (1981).
\bibitem{F2}
M.E. Fisher, in {\it{Critical Phenomena}}, ed. F.J. Hahne, Lecture
Notes in Physics, Vol. 186 (Springer, Berlin, 1983), p.1.
\bibitem{PF}
V. Privman and M.E. Fisher, J. Stat. Phys. {\bf{33}}, 385 (1983).
\bibitem{15}
X.S. Chen and V. Dohm, cond-mat/9711298.
\bibitem{EDC}
A. Esser, V. Dohm, and X.S. Chen, Physica {\bf{A 222}},
355 (1995); A. Esser, V. Dohm, M. Hermes, and J.S. Wang, Z. Phys. {\bf B97},
205 (1995).
\bibitem{CDS}
X.S. Chen, V. Dohm, and N. Schultka, Phys. Rev. Lett. {\bf{77}},
3641 (1996); X.S. Chen and V. Dohm, Physica {\bf{A 235}}, 555 (1997);
Int. J. Mod. Phys. {\bf{B 12}}, 1277 (1998).
\bibitem{CDT}
X.S. Chen, V. Dohm, and A.L. Talapov, Physica {\bf A 232}, 375 (1996).
\bibitem{LL}
L.D. Landau and E.M. Lifschitz, {\it{Statistical Physics}} (Pergamon Press,
London, 1959).
\bibitem{1}J. Rudnick, G. Gaspari, and V. Privman, Phys. Rev. {\bf B32}, 
7594 (1985).
\bibitem{sum} J. Shapiro and J. Rudnick, J. Stat. Phys. {\bf 43}, 51 (1986).
\bibitem{3} H.W. Diehl, in {\sl Phase Transitions and Critical Phenomena}, 
edited by C. Domb and J.L. Lebowitz (Academic, London, 1986), Vol. 10, p.76.
\bibitem{4}Y.Y. Goldschmidt, Nucl. Phys. {\bf B280 }, 340 (1987);
{\bf B285}, 519 (1987). 
\bibitem{5} J.C. Niel and J. Zinn-Justin, Nucl. Phys. {\bf B280}, 355 (1987).
\bibitem{6} H.W. Diehl, Z. Phys. {\bf B66 }, 211 (1987).
\bibitem{7} S. Singh and R.K. Pathria, Phys. Rev. {\bf B38}, 2740 (1988).
\bibitem{8} H.K. Janssen, B. Schaub, and B. Schmittmann, Z. Phys. {\bf B71 }, 
377 (1988); J. Phys. {\bf A 21 }, L427 (1988).
\bibitem{9} K. Binder and J.S. Wang, J. Stat. Phys. {\bf 55}, 87 (1989). 
\bibitem{10} K. Binder, in {\sl Finite Size Scaling and Numerical Simulation 
of Statistical Systems}, edited by V. Privman 
(World Scientific, Singapore, 1990), p.173.
\bibitem{a1}V. Privman, P.C. Hohenberg, and A. Aharony, in 
{\sl Phase Transitions and Critical Phenomena}, 
edited by C. Domb and J.L. Lebowitz (Academic, New York, 1991), Vol. 14, p.1.
\bibitem{a2} M. Krech, {\sl The Casimir Effect in Critical Systems} 
(World Scientific, Singapore, 1994).
\bibitem{a3} A. Aharony and D. Stauffer, Physica {\bf A215}, 242 (1995).
\bibitem{a4} H.K. Janssen and W. Koch, Physica {\bf A227}, 66 (1996).
\bibitem{a5} J. Cardy, {\sl Scaling and Renormalization in Statistical 
Physics} (Cambridge University Press, Cambridge, 1996).
\bibitem{a6} U. Ritschel and H.W. Diehl, Nucl. Phys. {\bf B464 }, 512 (1996).
\bibitem{a7} K. Binder, Rep. Prog. Phys. {\bf 60}, 487 (1997).
\end{thebibliography}
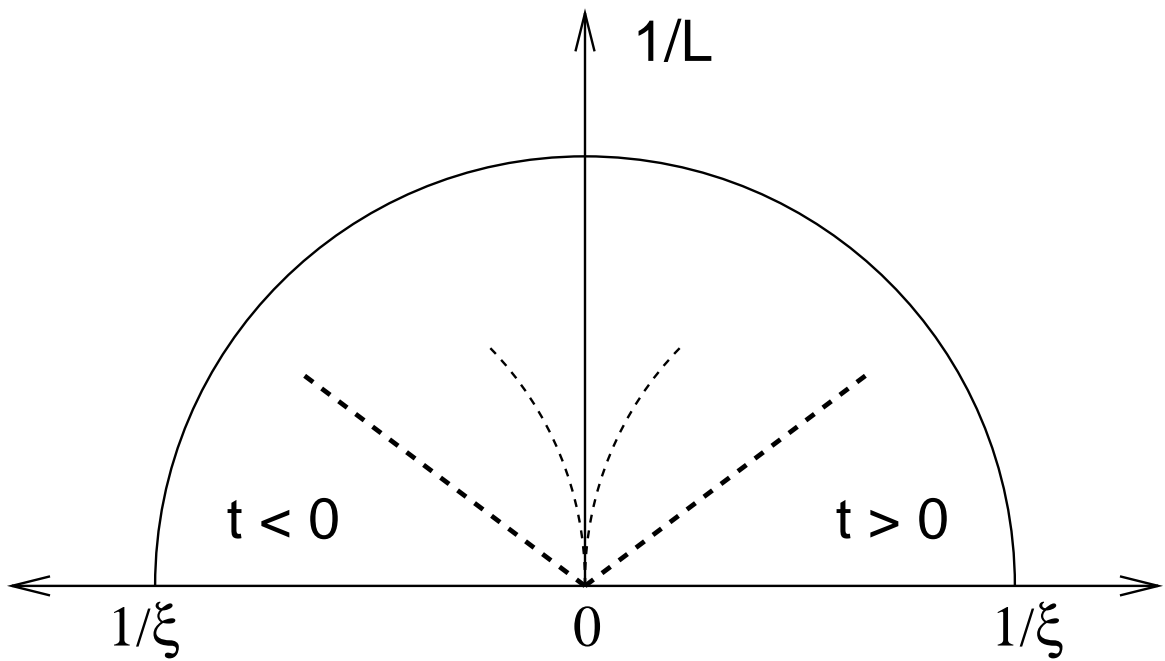
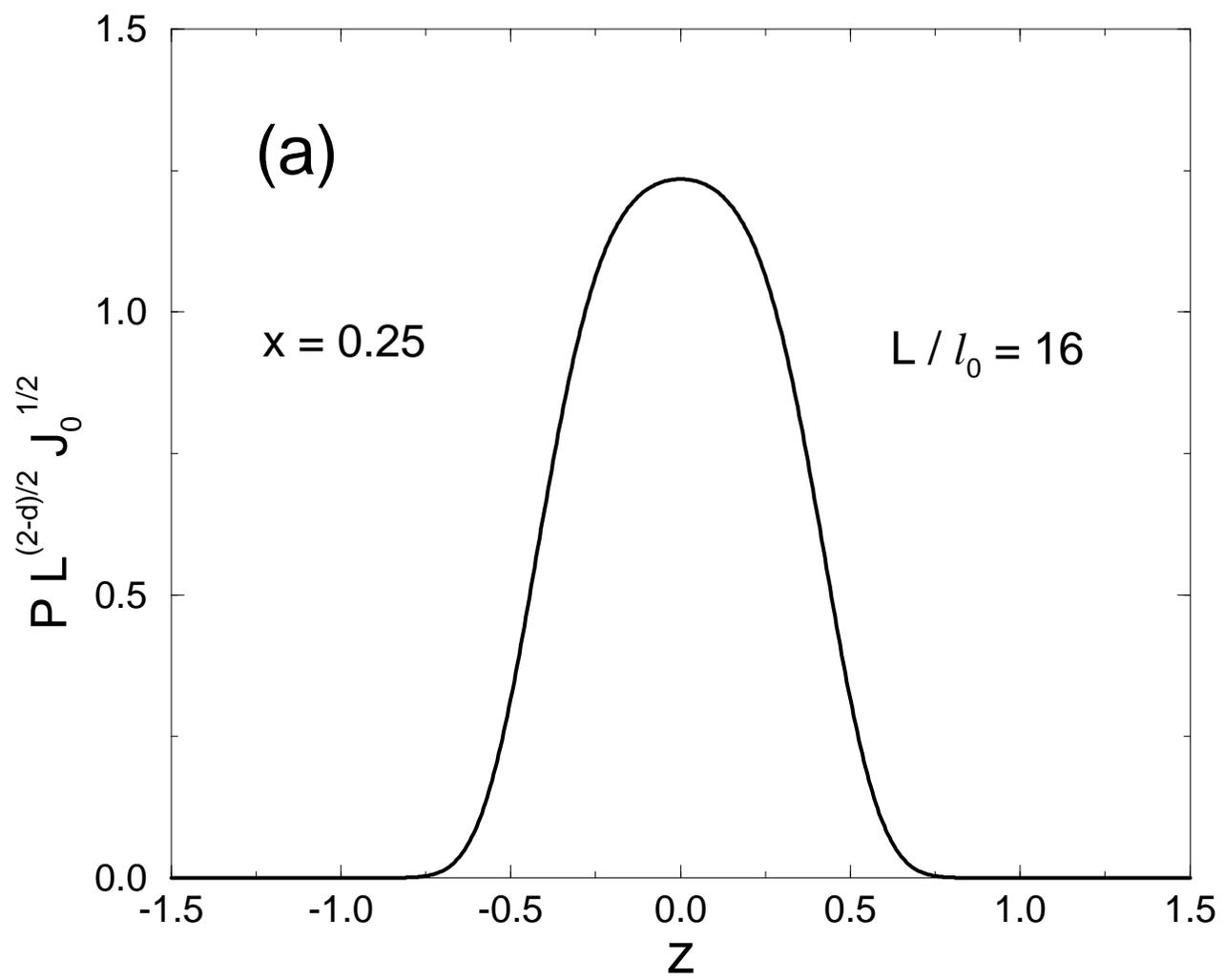
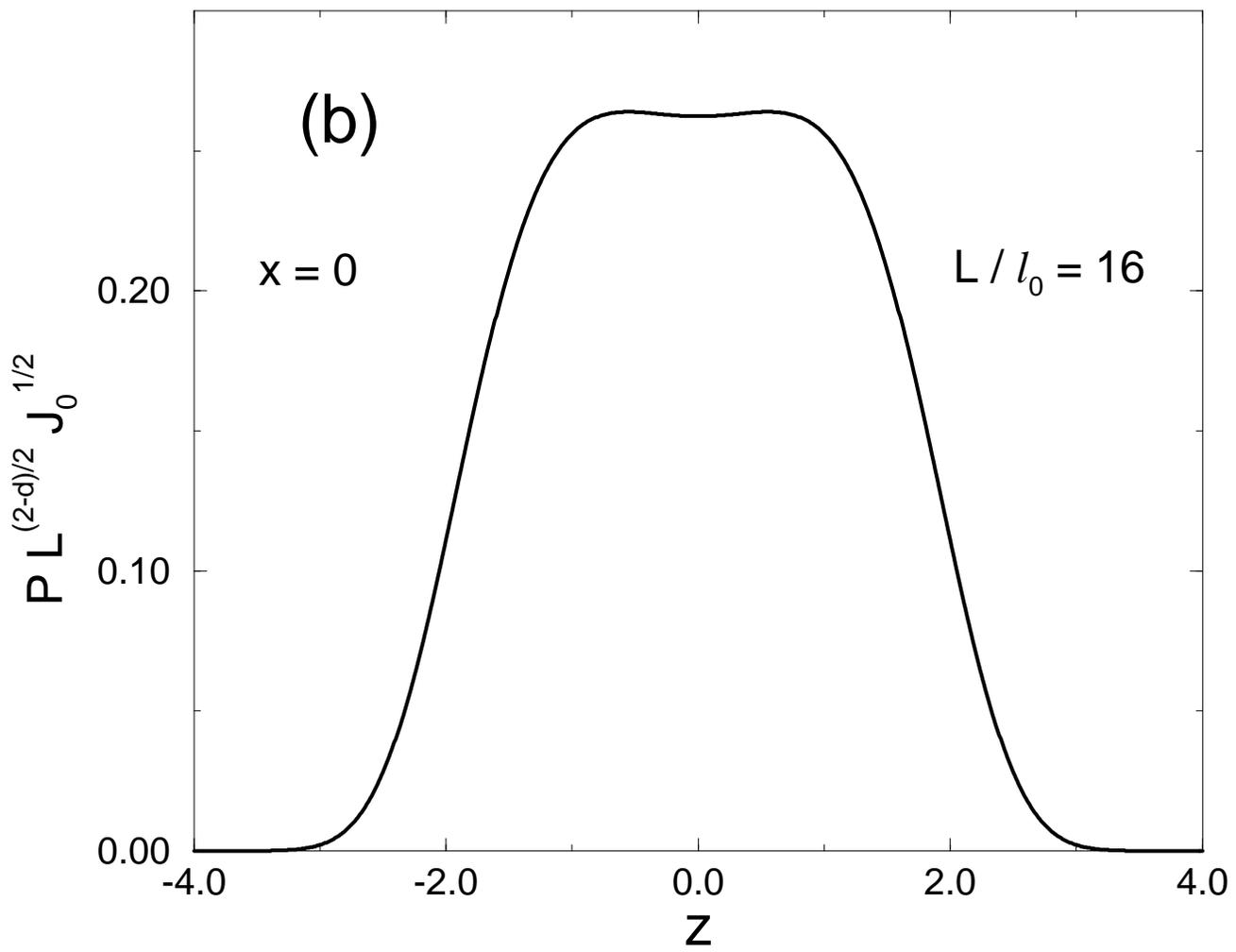
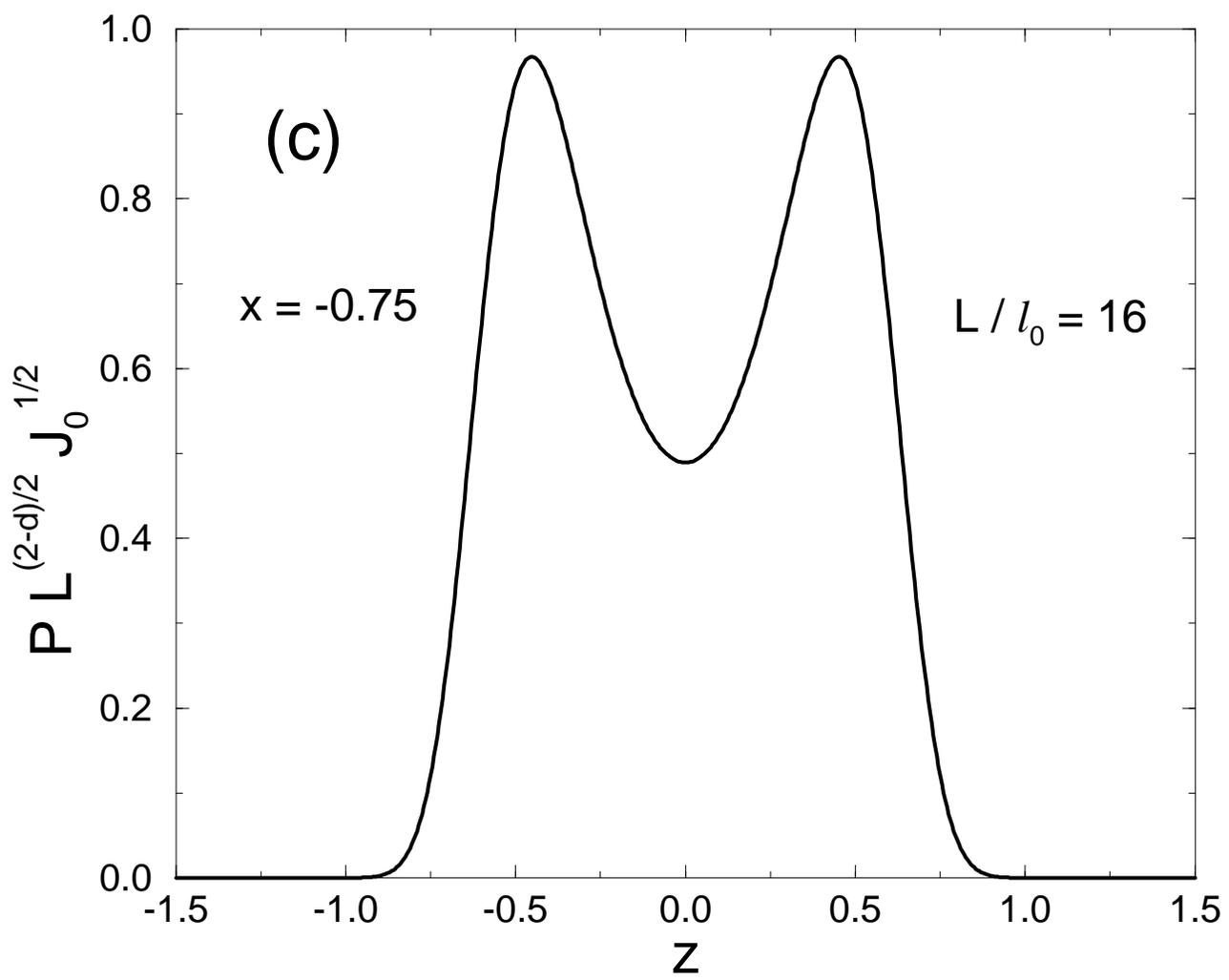
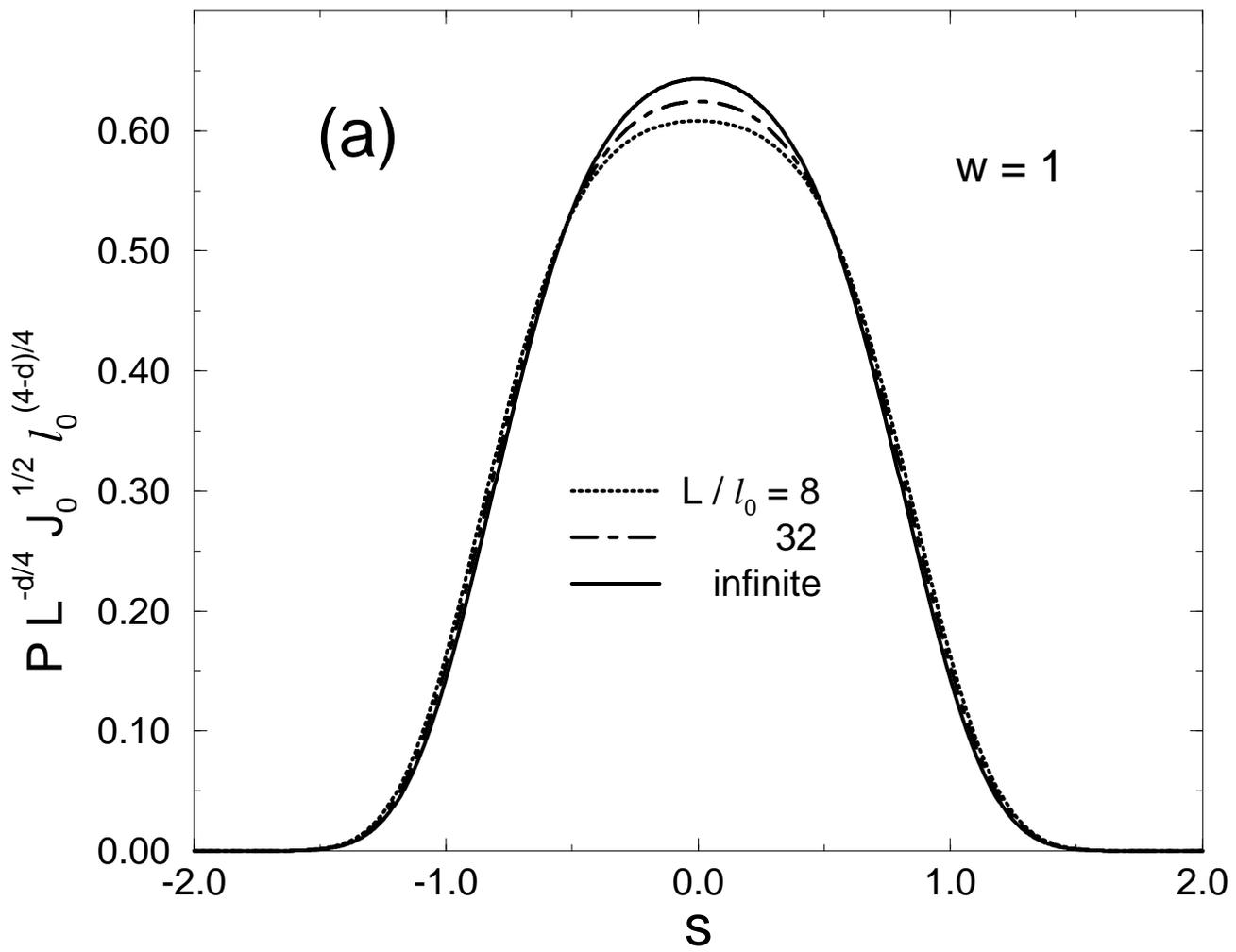
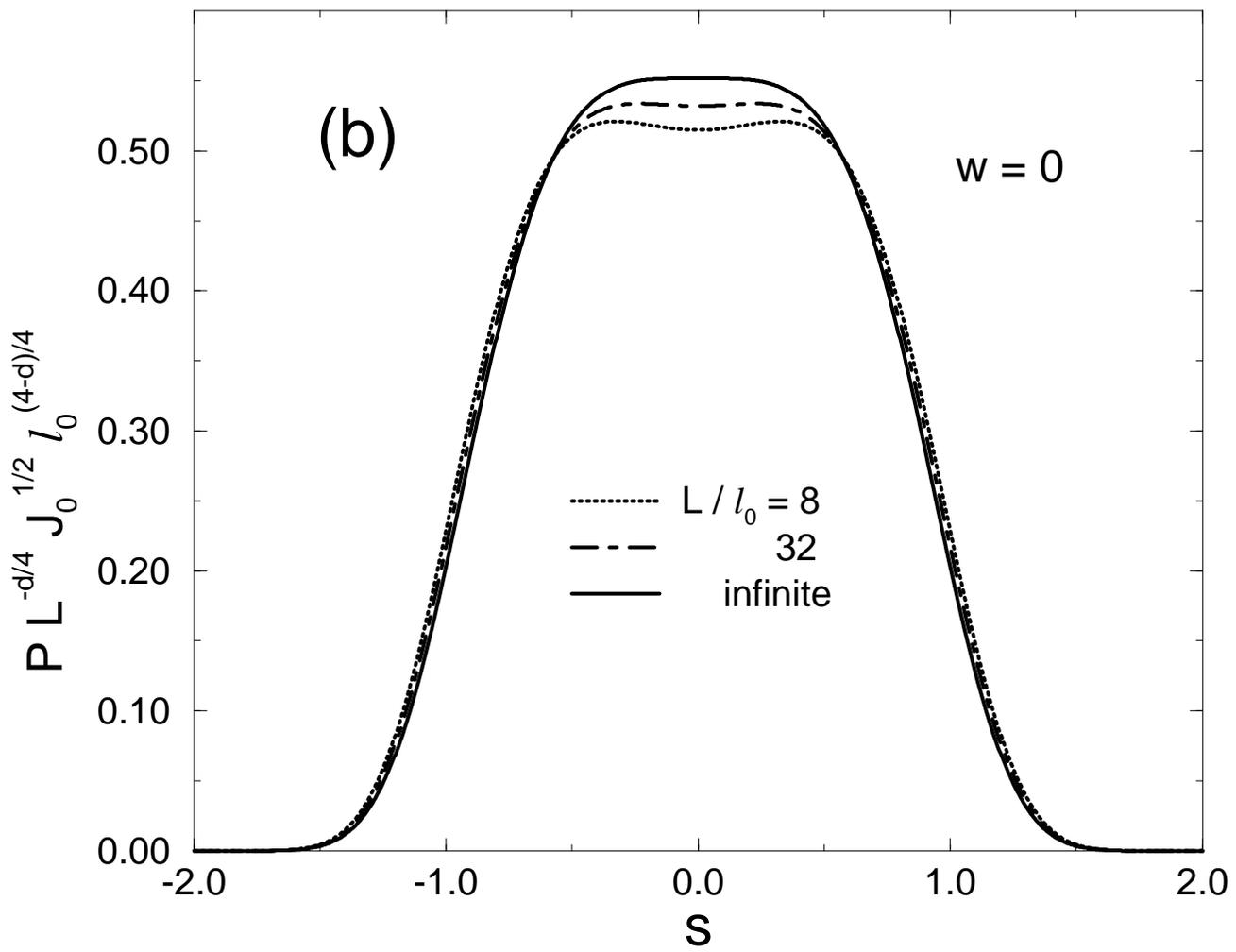
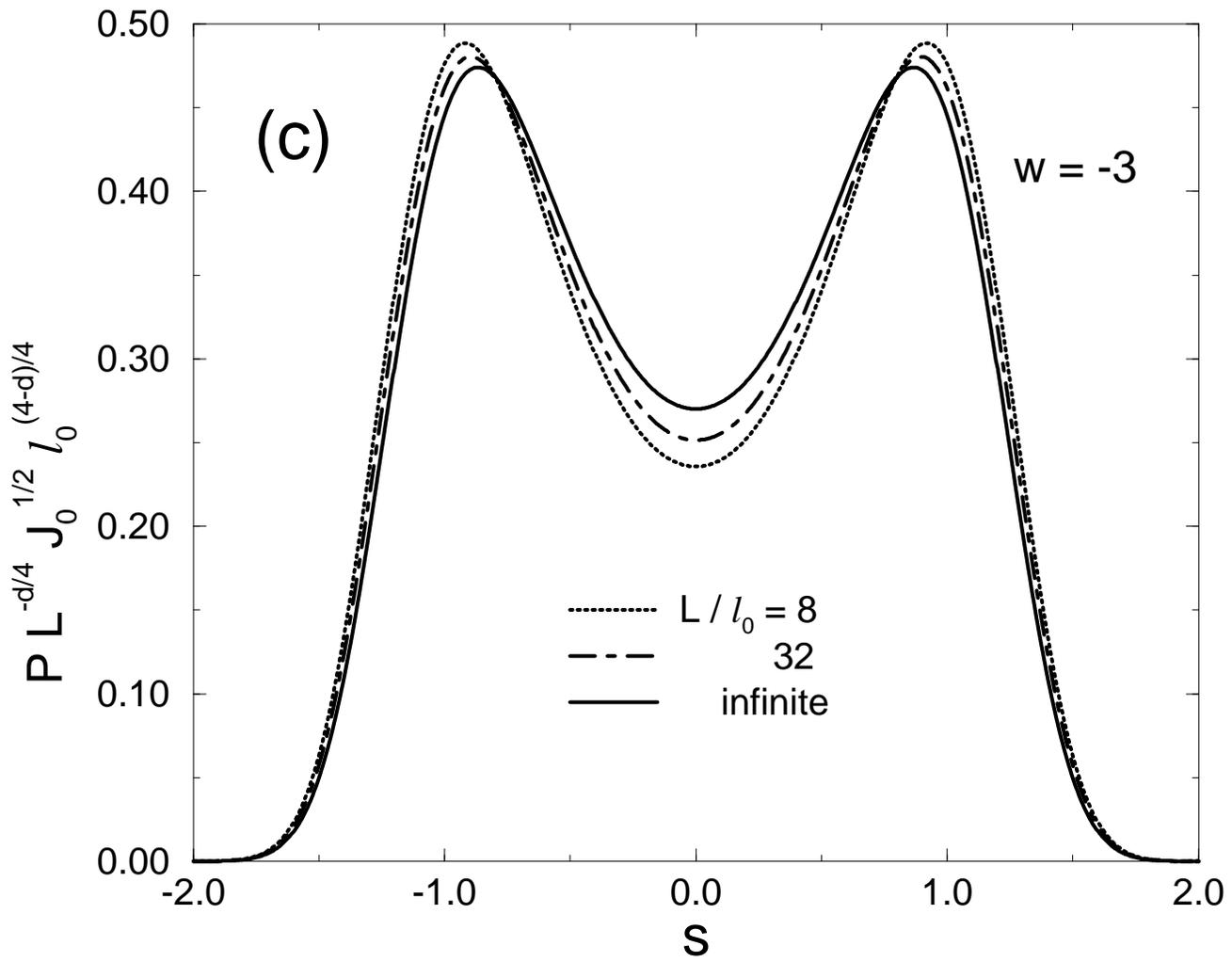
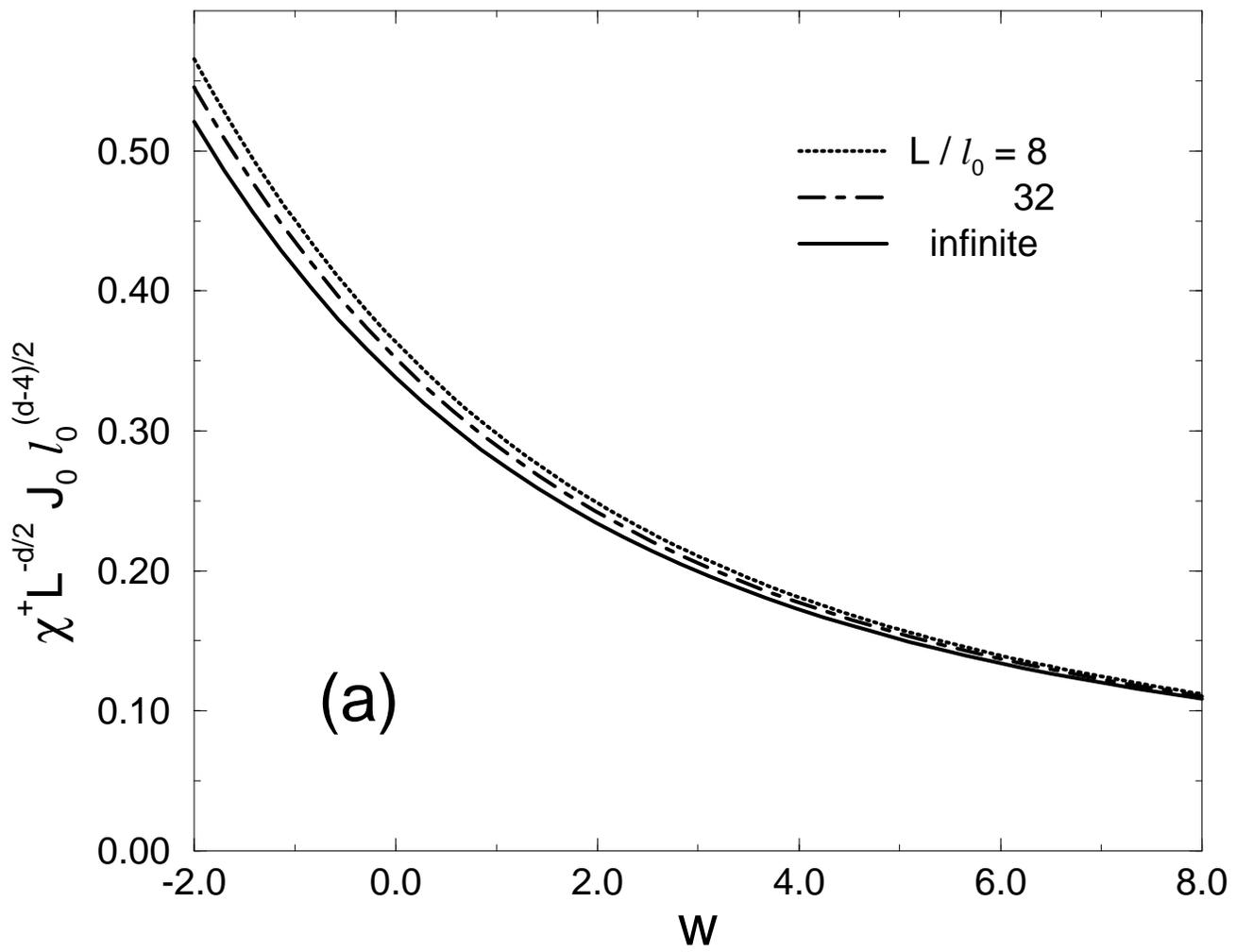
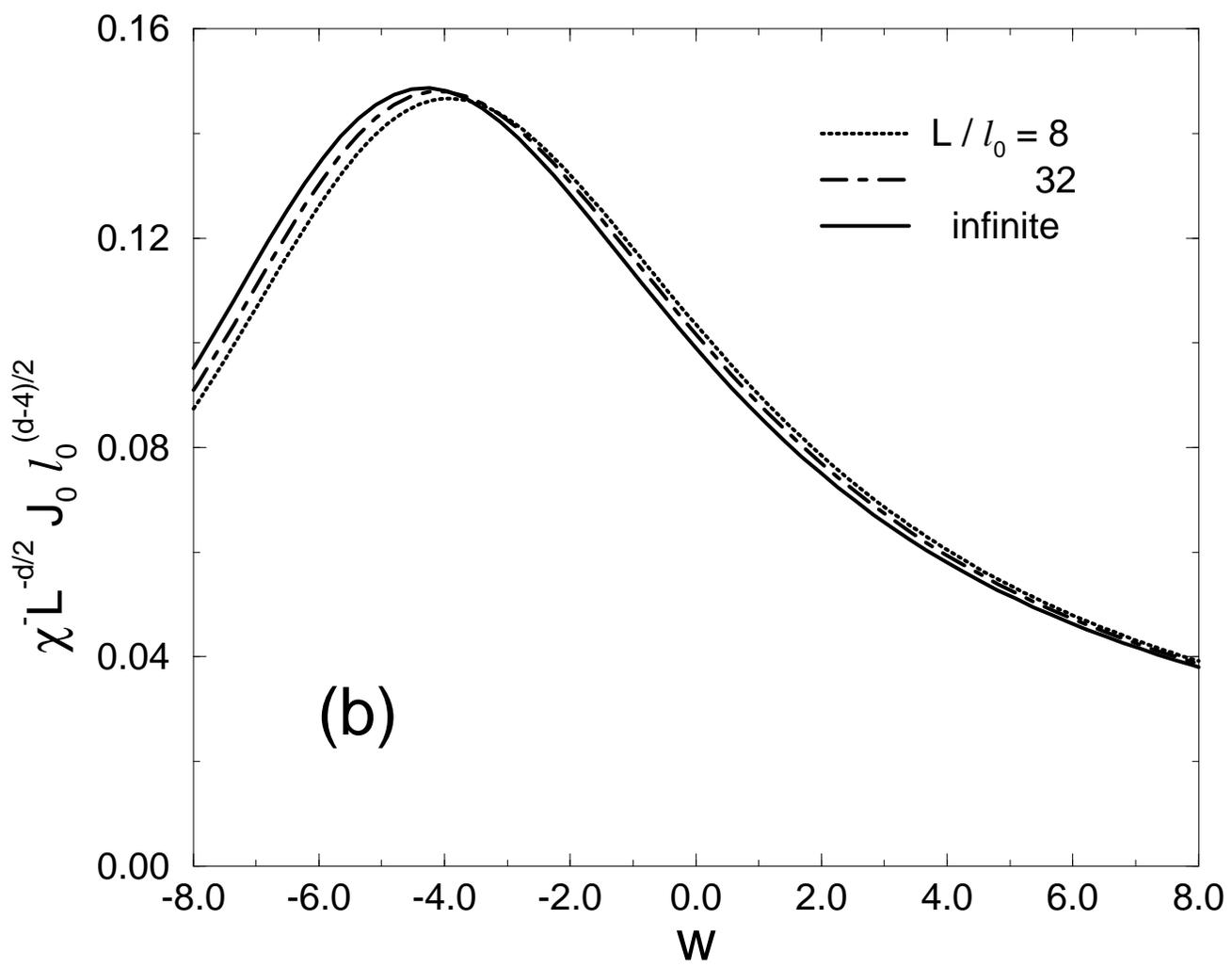
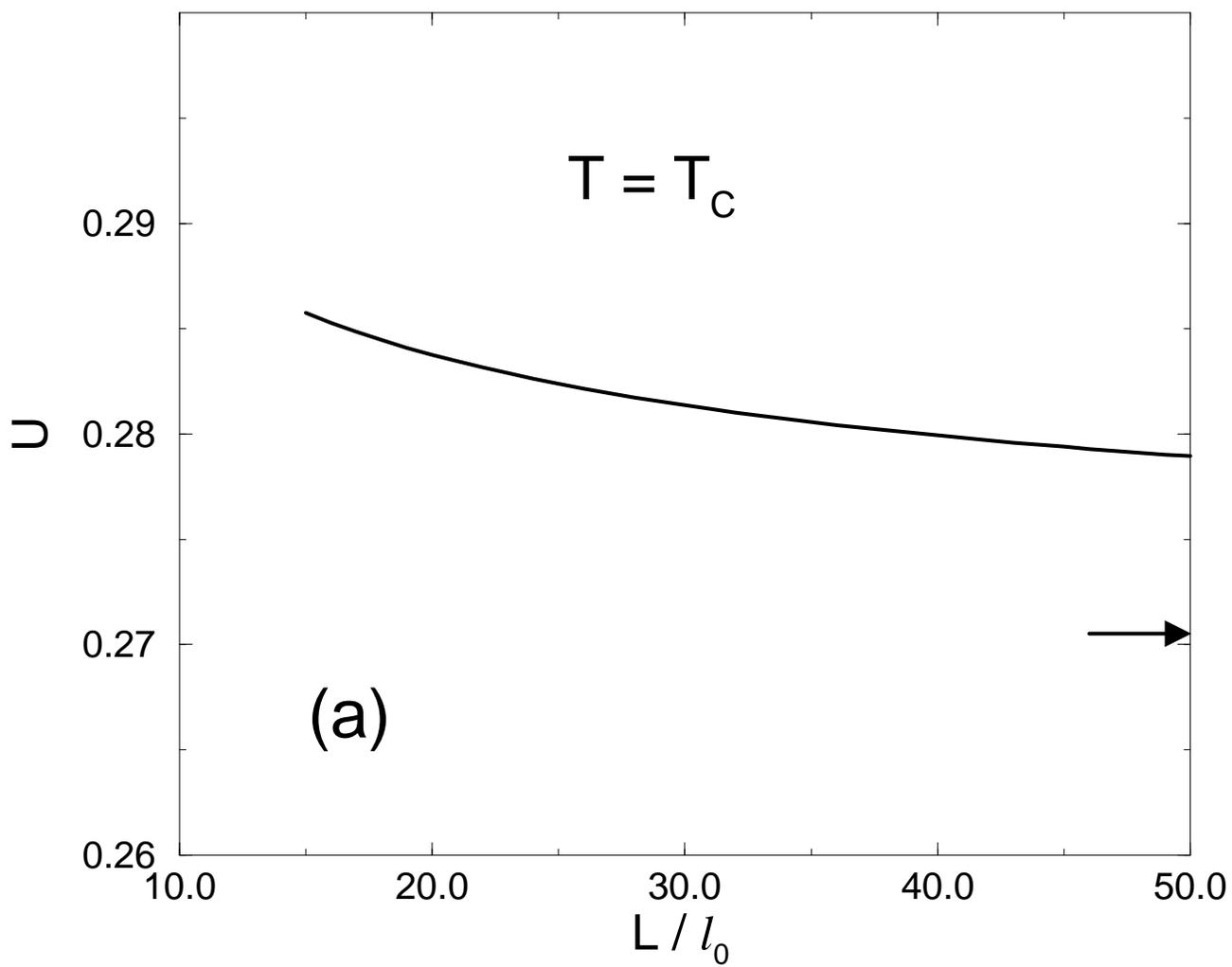
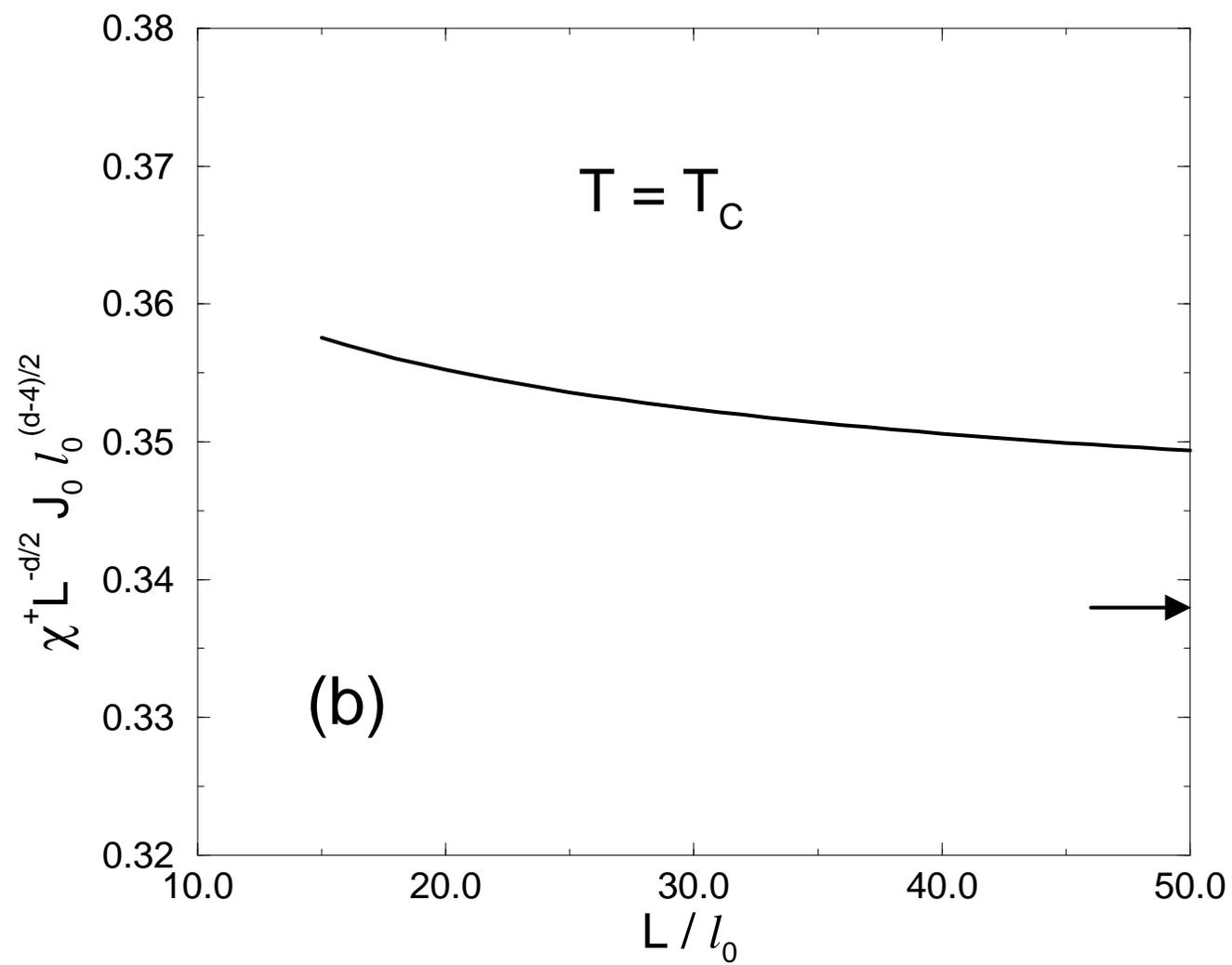
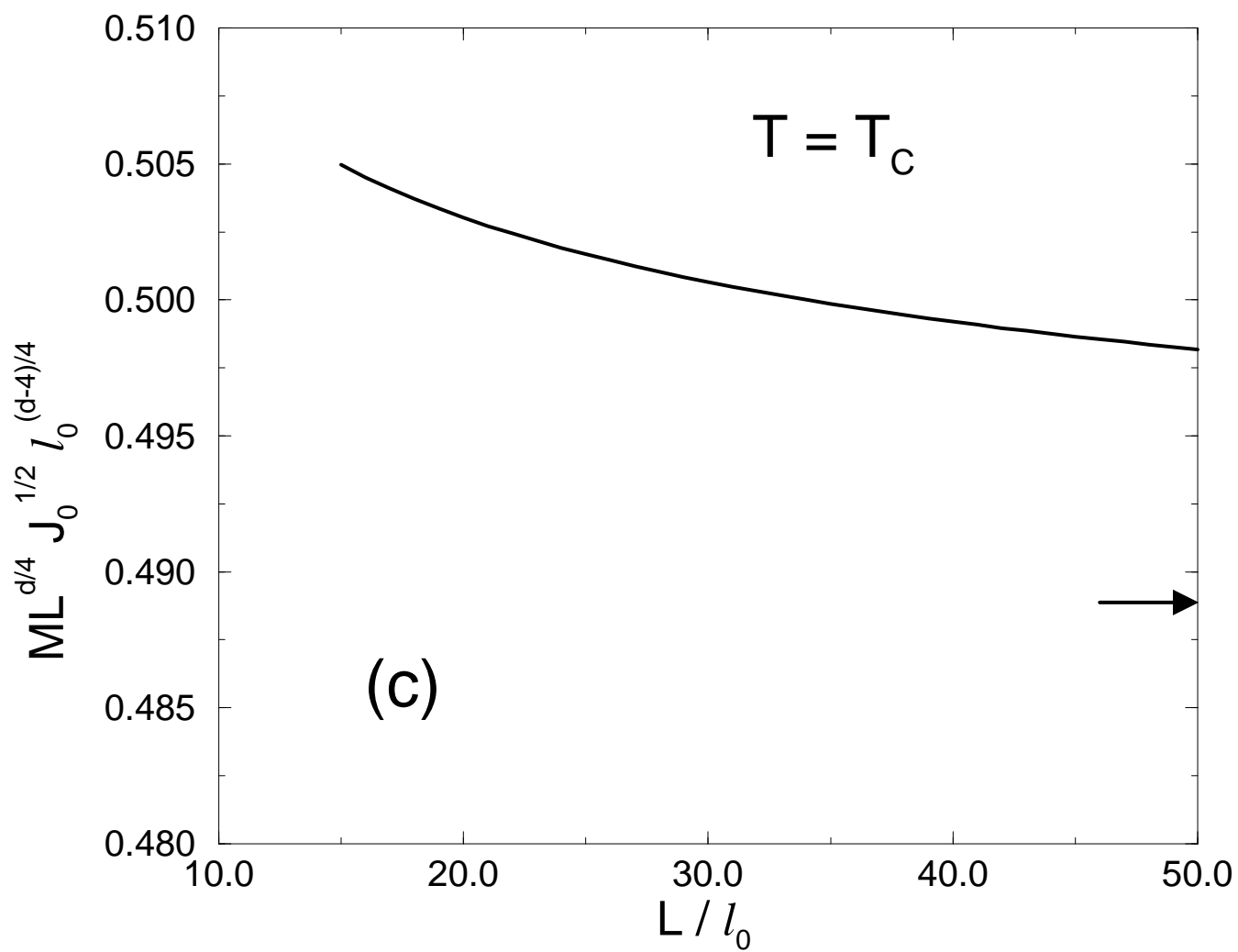
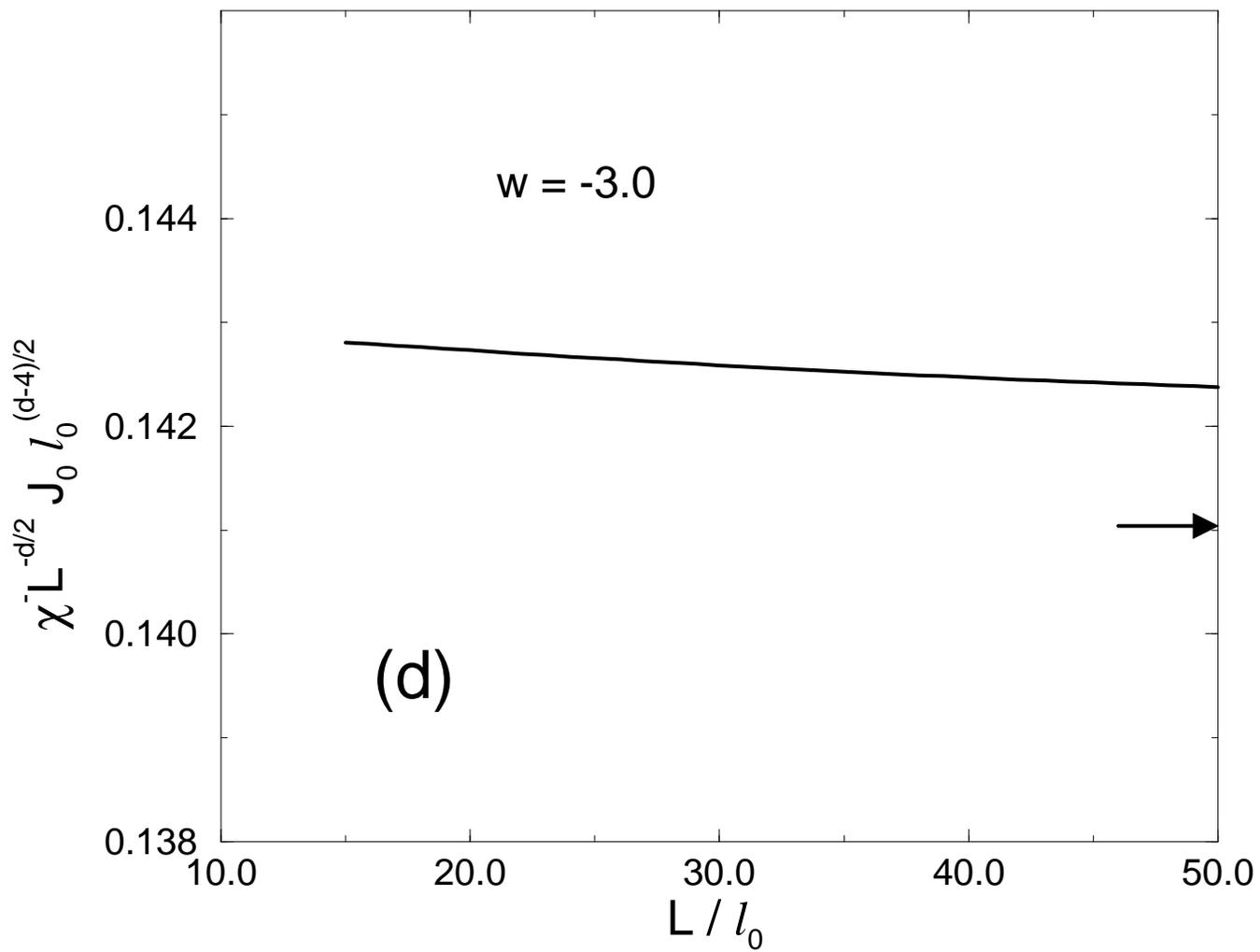
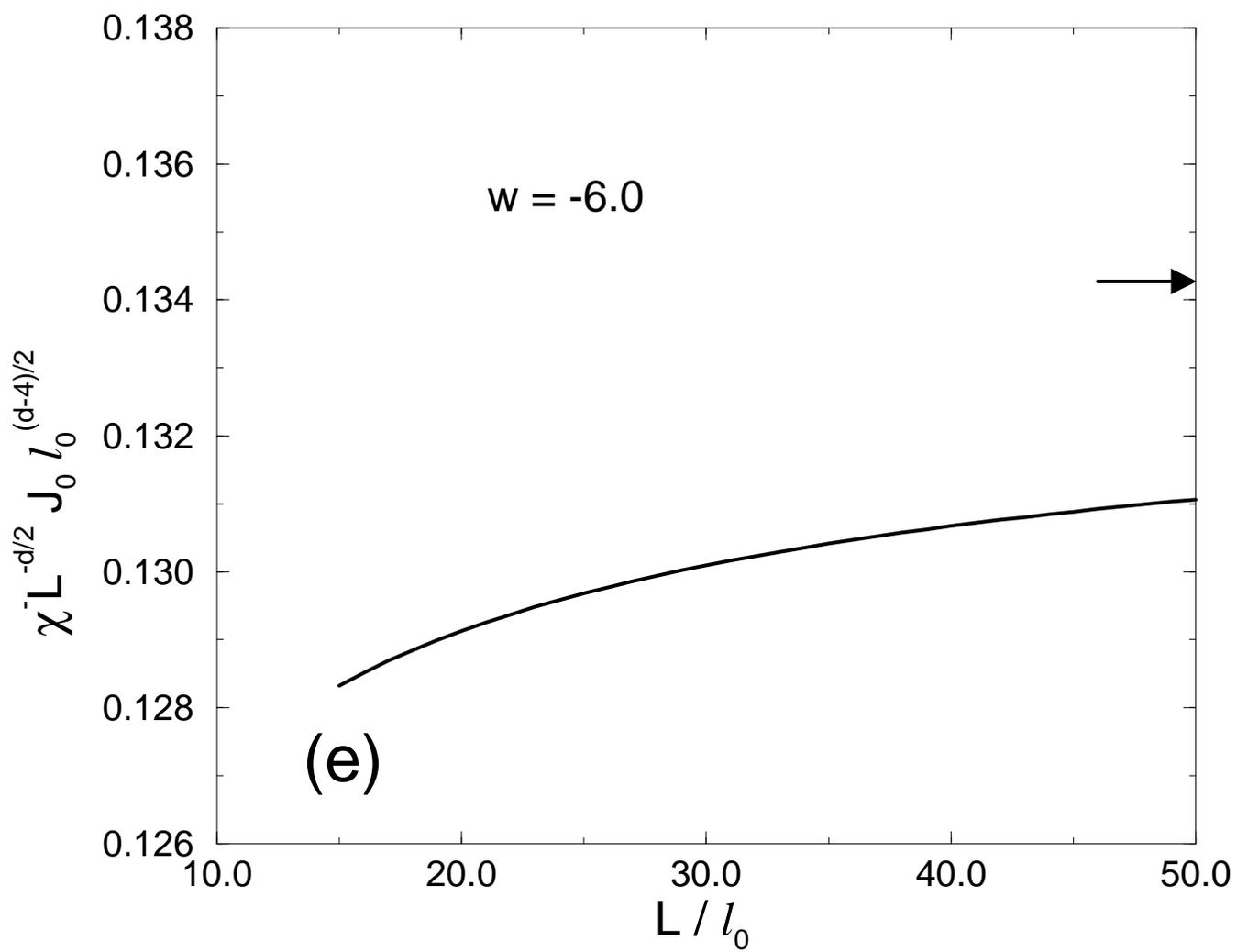
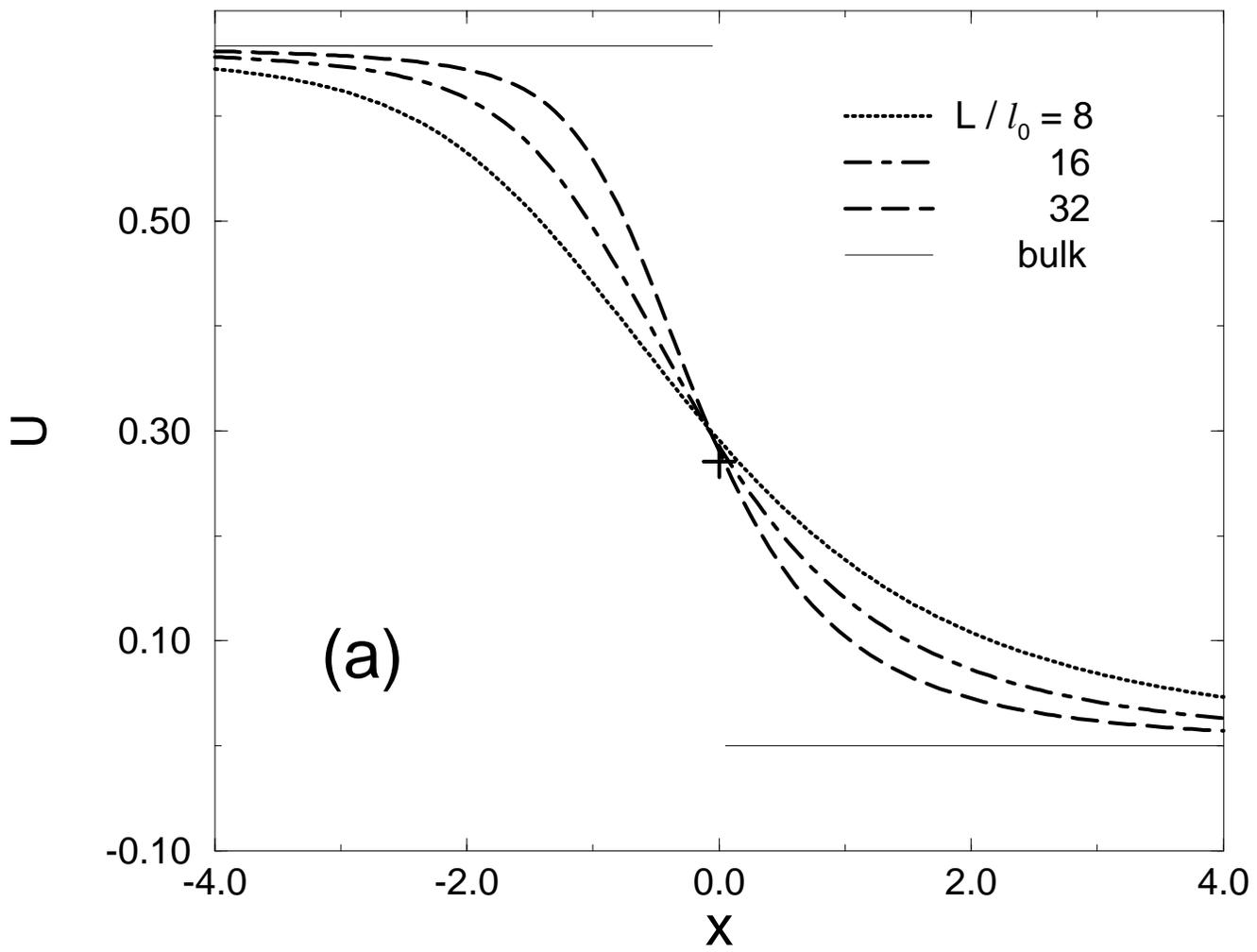
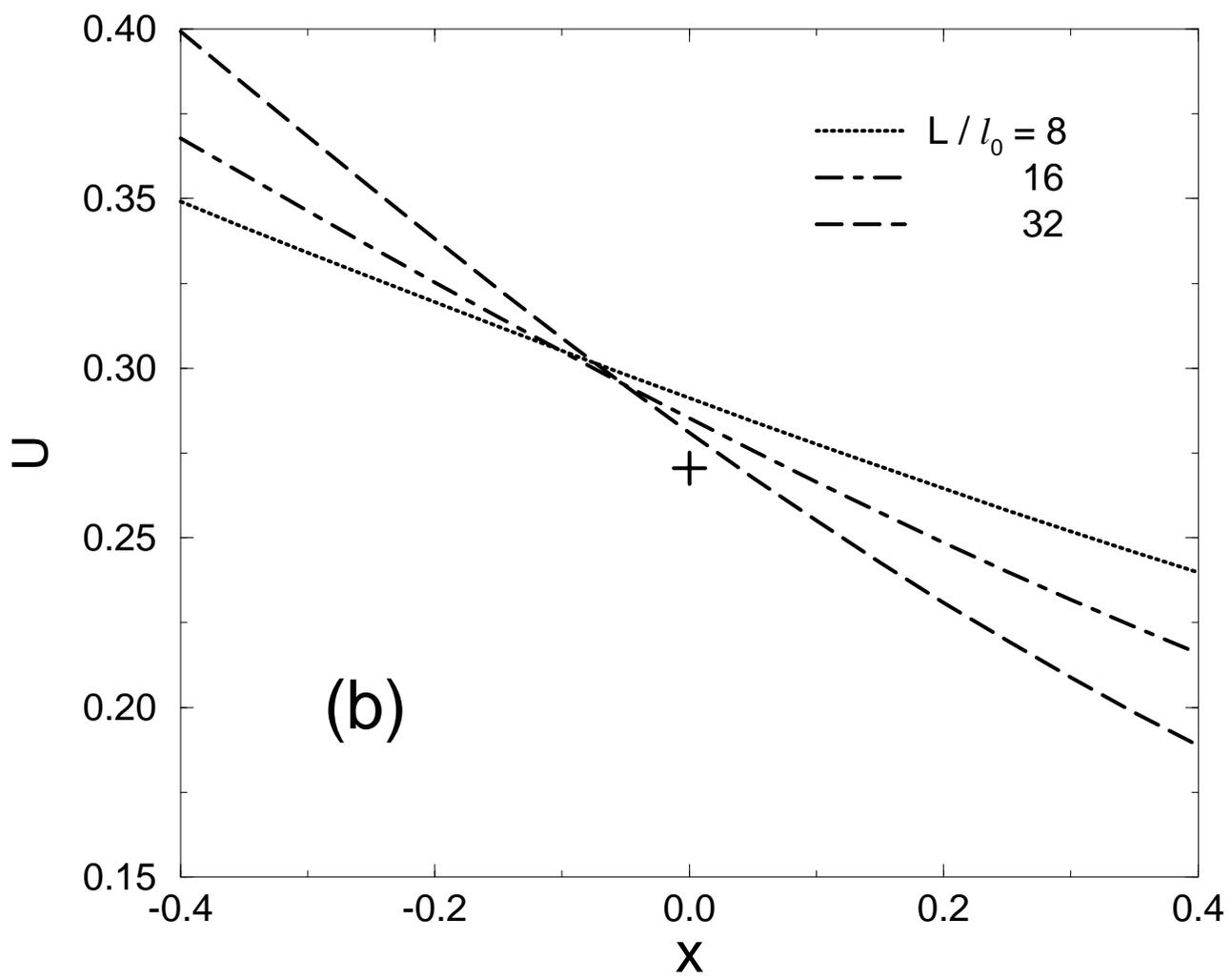
\end{document}